\documentclass[fleqn,usenatbib]{mnras}

\usepackage{natbib}
\usepackage{soul}

\usepackage[T1]{fontenc}
\usepackage{ae,aecompl}

\usepackage{outlines}
\usepackage{color}

\usepackage{graphicx}	
\usepackage{amsmath}	
\usepackage{amssymb}	
\usepackage{bm}	        
\usepackage{newtxtext,newtxmath}

\usepackage{natbib}
\def\xihm{\xi_{\mathrm{hm}}}
\def\ximm{\xi_{\mathrm{mm}}}
\def\mrm{\mathrm}
\defcitealias{Zu2021}{Paper I}
\newcommand{\lcdm}{\Lambda\mathrm{CDM}}
\newcommand{\avg}[1]{\left\langle #1 \right\rangle}

\newcommand{\dd}{\mathrm{d}}
\newcommand{\hmsol}{h^{-1}M_{\odot}}
\newcommand{\ms}{M_*}
\newcommand{\msbcg}{M_*^{\texttt{BCG}}}
\newcommand{\mh}{M_h}
\newcommand{\ptwod}{p(M_*^{\texttt{BCG}}, \lambda |M_h)}
\newcommand{\plntwod}{p(\ln M_*^{\texttt{BCG}}, \ln \lambda \mid M_h)}

\newcommand{\plntwodobs}{p(\ln M_*^{\texttt{BCG}}, \ln \lambda)}
\newcommand{\mpc}{\mathrm{Mpc}}
\newcommand{\hmpc}{h^{-1}\mathrm{Mpc}}

\newcommand{\hkpc}{h^{-1}\mathrm{kpc}}

\newcommand{\hhmsol}{h^{-2}M_{\odot}}
\newcommand{\ds}{\Delta\Sigma}

\newcommand{\siglam}{\sigma_{\ln\lambda\mid M_h}}
\newcommand{\sigms}{\sigma_{\ln M_*^{\texttt{BCG}}\mid M_h}}
\newcommand{\siglamknot}{\sigma_{\ln \lambda,0}}

\newcommand{\kms}{\mathrm{km}\,s^{-1}}

\newcommand\redmapper{redMaPPer}

\newcommand{\rom}[1]{\uppercase\expandafter{\romannumeral #1\relax}}
\title[Cluster Conformity and Assembly Bias]{Strong Conformity and Assembly
Bias: Towards a Physical Understanding of the Galaxy-Halo Connection in SDSS
Clusters}
\author[Zu 2021]{Ying  Zu$^{1, 2}$\thanks{E-mail: yingzu@sjtu.edu.cn},
Yunjia Song$^{1}$,
Zhiwei Shao$^{1}$,
Xiaokai Chen$^{1}$,
Yun Zheng$^{1}$,
Hongyu Gao$^{1}$,
\newauthor
Yu Yu$^{1, 2}$,
Huanyuan Shan$^{3}$,
Yipeng Jing$^{1, 2, 4}$
\\ \\
$^{1}$Department of Astronomy, School of Physics and Astronomy, Shanghai Jiao Tong
University, Shanghai 200240, China\\
$^{2}$Shanghai Key Laboratory for Particle Physics and Cosmology, Shanghai Jiao Tong
University, Shanghai 200240, China\\
$^{3}$Key Laboratory for Research in Galaxies and Cosmology, Shanghai Astronomical Observatory, Shanghai 200030, China \\
$^{4}$Tsung-Dao Lee Institute, Shanghai Jiao Tong University, Shanghai, 200240, China\\
}

\date{Accepted XXX. Received YYY; in original form ZZZ}

\pubyear{2021}

\begin{document}

\label{firstpage}
\pagerange{\pageref{firstpage}--\pageref{lastpage}}
\maketitle

\begin{abstract}
    Understanding the physical connection between cluster galaxies and
    massive haloes is key to mitigating systematic uncertainties in
    next-generation cluster cosmology. We develop a novel method to infer
    the level of conformity between the stellar mass of the bright central
    galaxies~(BCGs) $\msbcg$ and the satellite richness $\lambda$, defined
    as their correlation coefficient $\rho_{\mathrm{cc}}$ at fixed halo
    mass, using the abundance and weak lensing of SDSS clusters as
    functions of $\msbcg$ and $\lambda$. We detect a halo mass-dependent
    conformity as
    $\rho_{\mathrm{cc}}{=}0.60{+}0.08\ln(\mh/3{\times}10^{14}\hmsol)$.  The
    strong conformity successfully resolves the ``halo mass equality''
    conundrum discovered in Zu et al. (2021) --- when split by $\msbcg$ at
    fixed $\lambda$, the low and high-$\msbcg$ clusters have the same
    average halo mass despite having a $0.34$ dex discrepancy in average
    $\msbcg$. On top of the best--fitting conformity model, we develop a
    cluster assembly bias~(AB) prescription calibrated against the
    \texttt{CosmicGrowth} simulation, and build a conformity+AB model for
    the cluster weak lensing measurements. Our model predicts that with a
    ${\sim}20\%$ lower halo concentration $c$, the low-$\msbcg$ clusters
    are ${\sim}10\%$ more biased than the high-$\msbcg$ systems, in
    good agreement with the observations. We also show that the
    observed conformity and assembly bias are unlikely due to projection
    effects. Finally, we build a toy model to argue that while the
    early-time BCG-halo co-evolution drives the $\msbcg$-$c$ correlation,
    the late-time dry merger-induced BCG growth naturally produces the
    $\msbcg$-$\lambda$ conformity despite the well-known anti-correlation
    between $\lambda$ and $c$. Our method paves the path towards
    simultaneously constraining cosmology and cluster formation with future
    cluster surveys.
\end{abstract}




\vspace{1in}
\section{Introduction}
\label{sec:intro}

Collapsed from the highest peaks in the initial matter density field,
galaxy clusters are one of the most sensitive probes of cosmic
growth~\citep[see~\S 6 of][for a comprehensive review]{Weinberg2013}. With
the advent of all-sky optical imaging surveys, the abundance of clusters
with accurate halo mass measurements from weak gravitational lensing
provides stringent constraints on the matter density $\Omega_m$ and
clustering amplitude $\sigma_8$~\citep{Rozo2010, Zu2014, Abbott2020}, the
total mass of the neutrinos~\citep{Carbone2012, Costanzi2013,
Sartoris2016}, and the nature of gravity~\citep{Lam2012, Zu2014b,
Cataneo2018}.  However, cosmology with optical clusters requires a thorough
understanding of the connection between dark matter haloes and cluster
member galaxies, including both the satellite galaxies and the bright
central galaxies~(BCGs\footnote{We deliberately avoid the more
commonly-used nomenclature of ``brightest cluster galaxies'' as BCGs,
because we are interested in the properties of the central galaxies, which
are not necessarily the brightest members in their host
clusters~\citep{Chen2021}.}). In
this paper, we investigate the level of BCG-satellite conformity and
cluster assembly bias for a large sample of clusters observed by the Sloan
Digital Sky Survey~\citep[SDSS;][]{York2000}, in hopes of developing a
comprehensive model for interpreting the weak lensing of
clusters~\citep{Mandelbaum2018, Umetsu2020} in next-generation imaging
surveys.

The ``conformity'' phenomenon was originally detected by \citet{Weinmann2006}
inside the SDSS galaxy groups~\citep{Yang2007}. They found that the early-type
fraction of satellite galaxies is significantly higher in a halo with an
early-type central than in a halo {\it of the same mass} but with a late-type
central. Similar group-scale conformities were reported for neutral gas
fraction~\citep{Kauffmann2010}, emission features~\citep{Robotham2013}, and
quenching efficiency~\citep{Phillips2014, Knobel2015}. Such a conformity at
fixed halo mass suggests that a secondary halo property~\citep[e.g., halo
concentration;][]{Paranjape2015, Zu2018} affected the galaxy evolution within
clusters regardless of the central vs. satellite dichotomy. However, by applying
a group-finding algorithm to a conformity-free galaxy mock, \citet{Calderon2018}
demonstrated that the conformity signal could be spurious and likely entirely
caused by group-finding systematics.

For more massive systems, a conformity likely exists between the BCG stellar
mass~($\msbcg$) and the richness of massive satellite galaxies~($\lambda$), as
physical processes that tie the stellar mass growth of the BCGs to the BCG-satellite
interactions could naturally produce more massive BCGs in richer clusters at
fixed halo mass.  For instance, galactic cannibalism predicts that BCGs grow
primarily from dissipationless mergers with satellite galaxies that were
already in place at $z{=}2$~\citep{White1976, Ostriker1977}, and BCGs could also
grow their outskirts via the accretion of tidally disrupted
satellites~\citep{Wetzel2010}.  Observationally, \citet{Liu2009} found that the
fraction of BCGs in major dry mergers increases with the richness of the
clusters; \citet{To2020} inferred a positive correlation between the BCG
luminosity and $\lambda$ from analyzing the central and satellite luminosity
functions of SDSS clusters. However, They speculated that the correlation may be
induced by the projection effects~\citep{Zu2017, Busch2017, Costanzi2019b,
Sunayama2020, Myles2021, Grandis2021}, which could boost the estimated richness
for clusters in denser environments, hence earlier formation times and somewhat brighter
BCGs.

More recently, \citet[][hereafter referred to as~\citetalias{Zu2021}]{Zu2021}
measured the weak lensing signals $\ds$ for two subsamples of SDSS clusters,
split by $\msbcg$ at fixed $\lambda$. They discovered that the two subsamples
have equal average halo mass, despite having a ${\sim}0.34$ dex discrepancy in
$\msbcg$.  This apparent $\msbcg$-independence of halo mass is intriguing, as
models of cluster formation robustly predict that the average halo mass is a
increasing function of BCG stellar mass, with more massive BCGs generally
occupying haloes of higher mass. Therefore, for such a ``halo mass equality'' to
be observed between the low and high-$\msbcg$ clusters with the same $\lambda$
distribution, we expect a non-trivial correlation between $\msbcg$ and $\lambda$
{\it at fixed halo mass}, i.e., a conformity or anti-conformity between the BCG
and satellite galaxies.

Interestingly, \citetalias{Zu2021} also found that the high-$\msbcg$ clusters
have a higher average halo concentration and a lower halo bias, compared to
their low-$\msbcg$ counterparts with the same average halo mass. This
concentration--bias relation is potentially a detection of the ``cluster
assembly bias'' phenomenon, which was robustly predicted by $\lcdm$
simulations~\citep{Gao2005, Jing2007}. In principle, we can measure the average
halo mass and concentration from the small-scale $\ds$, as well as the average
halo bias from $\ds$ on large scales. However, previous studies focused
primarily on the measurement of halo mass from the small-scale $\ds$, while the
concentration-bias relation encoded in $\ds$ is largely unexplored due to their
relatively large measurement uncertainties from weak lensing. To extract unbiased
cosmological information from cluster weak lensing, it is imperative that we
incorporate the cluster assembly bias effect into the modelling of $\ds$
measurements from upcoming surveys with much smaller statistical uncertainties.

In this paper, we will firstly explore the ``halo mass equality'' conundrum
discovered in \citetalias{Zu2021} by explicitly modelling the correlation
between $\msbcg$ and $\lambda$ at fixed halo mass $\mh$, and then develop a
cluster assembly bias prescription for a simple yet comprehensive model of
cluster weak lensing. Our paper is accordingly organised into two main
parts.  In the first part of the paper, we describe the cluster catalogue,
BCG stellar mass estimates, and weak lensing measurements
in~\S\ref{sec:data}.  The statistical model of BCG--satellite conformity
and the Bayesian inference method are described in~\S\ref{sec:method}. We
present our model constraints and our solution to the ``halo mass
equality'' conundrum in ~\S\ref{sec:infer}. In the second part of the
paper, we develop a novel model for the cluster weak lensing by including
both conformity and assembly bias in~\S\ref{sec:dsfull}, supplemented by
the Appendices \S\ref{sec:dsfull_cab} and \S\ref{sec:dsfull_joint}. We
discuss the physical implications of our findings in~\S\ref{sec:physics}
and conclude by summarising our results and looking to the future in
~\S\ref{sec:conc}.

Throughout this paper, we assume the {\it Planck} cosmology~\citep{Planck2018}.
All the length and mass units in this paper are scaled as if the Hubble constant
is $100\,\kms\mpc^{-1}$. In particular, all the separations are co-moving
distances in units of $\hmpc$, and the halo and stellar mass are in units of
$\hmsol$ and $\hhmsol$, respectively. We adopt a spherical overdensity-based
halo definition so that the average halo density with the halo radius $r_{200m}$
is 200 times the mean density of the Universe, and the mass enclosed within
$r_{200m}$ is the halo mass $\mh$.  We use $\lg x{=}\log_{10} x$ for the
base-$10$ logarithm and $\ln x{=}\log_{e} x$ for the natural logarithm.

\section{Data}
\label{sec:data}

\subsection{Cluster Catalogue and Stellar Mass Estimates}
\label{subsec:data_cluster}

Following~\citetalias{Zu2021}, we employ the SDSS redMaPPer cluster
catalogue~\citep{Rykoff2014} derived by applying a red-sequence-based
photometric cluster finding algorithm to the SDSS DR8
imaging~\citep{Aihara2011}. Briefly, redMaPPer iteratively self-trains a model
of red-sequence galaxies calibrated by an input spectroscopic galaxy sample, and
then attempts to grow a galaxy cluster centred about every photometric
galaxy. Once a galaxy cluster has been identified by the matched-filters, the
algorithm iteratively solves for a photometric redshift based on the calibrated
red-sequence model, and re-centres the clusters about the best BCG candidates.

Therefore, each redMaPPer cluster is a conglomerate of red-sequence galaxies on
the sky, with each galaxy assigned a membership probability $p_{\mathrm{mem}}$
and a probability of being the BCG $p_{\mathrm{cen}}$.  For each cluster, the
richness $\lambda$ was computed by summing the $p_{\mathrm{mem}}$ of all member
galaxy candidates, and roughly corresponds to the number of red-sequence
satellite galaxies brighter than $0.2\,L_*$ within an aperture of
${\sim}1\,\hmpc$~(with a weak dependence on $\lambda$). At $\lambda{\geq}20$,
the SDSS \redmapper{} cluster catalogue is approximately volume-complete up to
$z{\simeq}0.33$, with cluster photometric redshift uncertainties as small as
$\delta(z)=0.006/(1+z)$~\citep{Rykoff2014, Rozo2015}.

We select $5476$ clusters with $\lambda{\geq}20$ and redshifts between
$0.17$ and $0.30$~($\avg{z}{=}0.242$) over a sky area of 10401 deg$^2$, and
pick the galaxy with the highest $p_{\mathrm{cen}}$ in each cluster as the
BCG. Among the $5476$ BCGs, $3610$ of them~(66 per cent) have
spectroscopic redshifts from SDSS, and for the $1866$ BCGs without
spectroscopy we assign them the photometric redshifts of their host
clusters.  We include $909$ more clusters than in
\citetalias{Zu2021}~($4567$ clusters), which excluded the area that was
masked out by the BOSS LOWZ galaxy sample~\citep{Dawson2013, Alam2015}.

Following~\citetalias{Zu2021}, we derive stellar masses for all BCGs by
fitting a two-component Simple Stellar Population~(SSP) template to their
extinction-corrected {\it gri} model magnitudes~(scaled to the $i$-band
$c$-model magnitudes). Following \citet{Maraston2009}, we assume the
dominant stellar population~(97 per cent) to be solar metallicity,
supplemented with a secondary~(3 per cent) metal-poor~($Z{=}0.008$)
population of the same age.  We utilize the \texttt{EzGal}
software~\citep{Mancone2012} and adopt the \citet{Bruzual2003} SSP model
and the \citet{Chabrier2003} IMF for the fits.  By examining the stacked
surface stellar mass density profiles of clusters at fixed $\msbcg$, we
infer the effective aperture of our $\msbcg$ estimates to be about
$35\,\hkpc$~\citep{Chen2021}. For a detailed comparison between
our photometric stellar mass estimates and the spectroscopic stellar masses
from ~\citet{Chen2012}, we refer interested readers to the Figure 1
in~\citetalias{Zu2021}.

However, there exists a systematic uncertainty in our central galaxy
stellar mass measurement due to the mis-centring effect, i.e., some of the
BCGs identified by the maximum $p_{\mathrm{cen}}$ are actually satellite
galaxies~\citep{Zhang2019}. From the weak lensing analysis,
\citetalias{Zu2021} inferred that ${\sim}30\%$ of the redMaPPer clusters in
our sample are mis-centred, and the mis-centring fraction decreases with
increasing $\msbcg$.  To assess the size of the systematic bias induced by
mis-centring, we examine the distribution of the stellar mass gaps
$\Delta\msbcg$ between galaxies with the maximum~(i.e. our BCG candidates)
and second highest $p_{\mathrm{cen}}$ in individual clusters.  We find that
in $25\%$ of the clusters the second probable central is more massive than
the BCG we select, and that among those clusters with $\Delta\msbcg{<}0$,
$70\%$ of them have $\Delta\msbcg{>}-0.1$ dex.  Therefore, assuming that
the mis-centred clusters are likely those with negative stellar mass gaps,
we expect that the BCG stellar mass of the mis-centred clusters could be
systematically underestimated by ${\sim}0.05{-}0.1$ dex.

\subsection{Cluster Weak Lensing Measurements}
\label{subsec:data_clusterwl}

\begin{table}
\renewcommand*{\arraystretch}{1.25}
    \centering \caption{Weak lensing mass estimates~(and associated
    uncertainties) of the redMaPPer clusters binned by $\lambda$, derived from
    \citet{Simet2017}. We assume 50 per cent of the uncertainties are systematic
    errors.}
\begin{tabular}{ccccc}
\hline
\hline
    $\lambda$ & [20,30) & [30,40) & [40,55) & [55, 100) \\
\hline
    $\lg\,M_h$ & $14.05{\pm}0.05$ & $14.25{\pm}0.05$ & $14.43{\pm}0.05$ &
    $14.64{\pm}0.05$ \\
\hline
\end{tabular}
\label{tab:wlmass}
\end{table}

We employ two sets of cluster weak lensing measurements in our analysis. For the
Bayesian analysis in~\S\ref{sec:infer}, we derive constraints on $\ptwod$, the
2D probability density function~(PDF) of the $\msbcg$ and $\lambda$ of clusters
at fixed $\mh$, by making use of the weak lensing halo mass measurements of
clusters in bins of $\lambda$ from ~\citet{Simet2017}. In particular, we assume
the best--fitting mass--richness relation inferred by~\citet{Simet2017}~(their
Equation 28), and compute the mean halo mass in each of the four richness bins,
which is listed in Table~\ref{tab:wlmass}. Following
\citet{Simet2017}~\citep[also see][]{Costanzi2019}, we assign 50 per cent of the
uncertainties as systematic errors, which we assume to be fully correlated
between different richness bins.  \citet{Murata2018} showed that the
mass--richness relation of \citet{Simet2017} derived from the SDSS imaging is
consistent with the recent measurements from the Hyper
Suprime-Cam~\citep[HSC;][]{Aihara2018, Mandelbaum2018b}.  We refer interested
readers to \citet{Simet2017} for technical details of the halo mass
measurements.

For testing whether our best--fitting models of $\ptwod$, in combination with
the cluster assembly bias prescription, can resolve the ``halo mass equality''
conundrum, we predict the surface density contrast profiles $\ds(r_p)$ for the
low and high-$\msbcg$ cluster subsamples, and compare to the weak lensing
measurements of the two subsamples made in~\citetalias{Zu2021} from the DECaLS
imaging~\citep{Dey2019}.  We will directly present the comparison
in~\S\ref{sec:dsfull} and refer readers to~\citetalias{Zu2021} for technical
details of the weak lensing measurements from DECaLS.

Note that we do not include the halo mass estimates for the low and
high-$\msbcg$ clusters from \citetalias{Zu2021} in our Bayesian analysis
of~\S\ref{sec:infer}, because the estimates from \citetalias{Zu2021} do not
include some of the systematic uncertainties considered by \citet{Simet2017},
including the shear calibration errors, photo-z biases, halo triaxiality, etc. Therefore, to avoid inhomogeneity in our input data, we only include the weak lensing halo mass in bins of $\lambda$ measured by
\citet{Simet2017} in our Bayesian analysis, but directly model the $\ds$
measurements from \citetalias{Zu2021} in \S\ref{sec:dsfull}.

\section{Methodology}
\label{sec:method}

The data vector of our Bayesian analysis in~\S\ref{sec:infer} consists of three
components,
\begin{itemize}
    \item $N_{\mathrm{cls}}=5476$: the total number of clusters observed with
        $\lambda{\geq}20$ and $0.17<z<0.30$ over a sky area of 10401 deg$^2$.
    \item $\{\msbcg, \lambda\}_{i=1 \cdots 5476}$: BCG stellar mass and
    satellite richness of the observed $5476$ {\it individual} clusters.
    \item  $\{ M_h \mid [\lambda_{\mathrm{min}}^j,
    \lambda_{\mathrm{max}}^j]\}_{j=1\cdots 4}$: Weak lensing halo mass
    measurements of four richness bins listed in Table~\ref{tab:wlmass}.
\end{itemize}
Below we will describe our analytic model for predicting each of the three
components.

\subsection{Modelling the 2D PDF of \texorpdfstring{$\bm{\msbcg}$}{Mstar} and \texorpdfstring{$\bm{\lambda}$}{lambda} at fixed \texorpdfstring{$\bm{M_h}$}{Mh}}
\label{subsec:method_ptwod}

The 2D PDF of $\msbcg$ and $\lambda$ at fixed $\mh$, $\ptwod$, is the centrepiece of our statistical model of galaxy-halo connection for clusters. Our model of $\ptwod$ consists of three components, the richness-to-halo mass
relation~(RHMR) that describes the 1D log-normal PDF of richness at fixed halo
mass $p(\lambda\mid \mh)$, the stellar-to-halo mass relation~(SHMR) that
specifies the 1D log-normal PDF of BCG stellar mass at fixed halo mass $p(\msbcg
\mid \mh)$, and the correlation coefficient between $\msbcg$ and $\lambda$ as a
function of halo mass $\rho_{\mathrm{cc}}(\mh)$. We will refer to models with
$\rho_{\mathrm{cc}}{>}0$ as ``conformity'' models and those with
$\rho_{\mathrm{cc}}{<}0$ as ``anti-conformity'' models, respectively.

We assume the mean RHMR to be
\begin{equation}
    \langle \ln\,\lambda \mid \mh \rangle = A + \alpha \ln
    \left(\frac{\mh}{M_{\mathrm{pivot}}}\right),
    \label{eqn:rhmr}
\end{equation}
where $A$ and $\alpha$ are the amplitude and slope of the power-law,
respectively, and we set the pivot halo mass $M_{\mathrm{pivot}}{=}3\times
10^{14}\hmsol$. Following \citet{Murata2018}, we further assume a
mass-dependent logarithmic scatter about the median RHMR,
\begin{equation}
    \siglam = \siglamknot + q \ln \left(\frac{\mh}{M_{\mathrm{pivot}}}\right),
    \label{eqn:siglam}
\end{equation}
where $\siglamknot$ is the scatter at $M_{\mathrm{pivot}}$, and $q$ is the slope
of the halo mass dependence. The combination of Equations~\ref{eqn:rhmr} and
\ref{eqn:siglam} thus fully describes the RHMR
\begin{equation}
p(\ln\,\lambda \mid \mh)\sim
    \mathcal{N}\left(\langle \ln\,\lambda \mid \mh \rangle,\,\siglam^{2}\right).
    \label{eqn:p1dlam}
\end{equation}

For the mean SHMR, we adopt a functional form proposed by \citet{Behroozi2010}
via its inverse function,
\begin{equation}
    \mh = M_1  m^\beta 10^{\left(m^\delta / (1 + m^{-\gamma}) - 1/2\right)},
    \label{eqn:shmr}
\end{equation}
where $m\equiv\ms/M_{*,0}$. Among the five parameters in
Equation~\ref{eqn:shmr}, $M_{h,1}$ and $M_{*,0}$ are the characteristic halo
mass and stellar mass that separate the behaviours in the low and high mass
ends. The inverse function starts with a low-mass end slope $\beta$, crosses a
transitional regime around ($M_{*,0}$, $M_{h,1}$) dictated by $\gamma$, and
reaches a high-mass end slope $\beta+\delta$. We assume a constant log-normal
scatter $\sigms$, because we are primarily interested in the massive end of the
SHMR, where the halo mass dependence of scatter was found to be
weak~\citep{Zu2015}.  Similarly, the combination of Equations~\ref{eqn:shmr} and
a constant scatter fully specifies the SHMR
\begin{equation}
p(\ln\,\msbcg \mid \mh)\sim
\mathcal{N}\left(\langle \ln\,\msbcg \mid \mh \rangle,\,\sigms^{2}\right).
    \label{eqn:p1dms}
\end{equation}

As mentioned in~\S\ref{sec:intro},~\citetalias{Zu2021} discovered that the
scatter of the SHMR is at least partially driven by the concentration of dark
matter haloes, so that the more massive BCGs are preferentially hosted by the
more concentrated haloes at fixed halo mass. Therefore, to accurately predict
the weak lensing profiles of clusters binned by $\msbcg$, we also need to take
into account the concentration--bias relation predicted by the halo assembly
bias effect, as will be discussed later in~\S\ref{sec:dsfull}.

To derive the joint 2D PDF $\ptwod$, we need to combine
Equation~\ref{eqn:p1dlam} and \ref{eqn:p1dms} into a bivariate Gaussian at each
halo mass via the correlation coefficient $\rho_{\mathrm{cc}}$ at that mass. To allow the
level of conformity between BCG and satellites to vary with halo mass, we assume
a halo mass dependence of $\rho_{\mathrm{cc}}$ as
\begin{equation}
    \rho_{\mathrm{cc}}(\mh) = \rho_{\mathrm{cc},0} + s \ln
    \left(\frac{\mh}{M_{\mathrm{pivot}}}\right), \label{eqn:rhocc}
\end{equation}
where $\rho_{cc,0}$ is the correlation coefficient at $M_{\mathrm{pivot}}$, and
$s$ describes the slope of the halo mass dependence.  Given the two mean
scaling relations and their associated scatters, it is now trivial to write out
the bivariate Gaussian form for $\plntwod$ as
\begin{equation}
    \plntwod =
    \frac{\exp\left(-(\hat{l}^2-2\rho_{\mathrm{cc}}\hat{l}\hat{m}_*+\hat{m}_*^2)/(2(1-\rho_{\mathrm{cc}}^2))\right)}{2\pi
    \siglam \sigms\sqrt{1-\rho^2_{cc}}},
    \label{eqn:plntwod}
\end{equation}
where $\hat{m}_*$ is the {\it relative} BCG stellar mass
\begin{equation}
    \hat{m}_* \equiv \frac{\ln\msbcg - \avg{\ln\msbcg\mid \mh}}{\sigms},
    \label{eqn:mhat}
\end{equation}
and $\hat{\lambda}$ the {\it relative} richness
\begin{equation}
    \hat{\lambda} \equiv \frac{\ln\lambda - \avg{\ln\lambda\mid \mh}}{\siglam}.
    \label{eqn:lhat}
\end{equation}

\subsection{Predicting Observing Probability of Each Cluster}
\label{subsec:method_pobs}

To predict the probability of observing any cluster with BCG mass $\msbcg$ and
satellite richness $\lambda$, we integrate $\ptwod$ over the halo mass function to obtain
\begin{equation}
    \plntwodobs = \frac{1}{n_{0}}\int_{M_h^{\mathrm{min}}}^{M_h^{\mathrm{max}}}  \plntwod
    \frac{\dd n}{\dd\mh} \dd\mh,
    \label{eqn:plntwodobs}
\end{equation}
where $\dd n/\dd\mh$ is the halo mass function at {\it Planck} cosmology, and
$n_0$ is the total number density of haloes between $M_h^{\mathrm{min}}$ and
$M_h^{\mathrm{max}}$. We choose the two integration limits to be $5\times
10^{11}\hmsol$ and $10^{16}\hmsol$, respectively, and adopt the
\citet{Tinker2008} fitting formula for $\dd n/\dd\mh$. With the integration
limits~(hence $n_0$) fixed, we can set $p(\mh){\equiv}\dd n/\dd\mh/n_0$ for the
rest of the paper. Note that although we evaluate all the quantities at each of the six
equal-width redshift slices between $z{=}0.17$ and $0.30$, and integrate over
the redshift range (with cluster photo-z uncertainty included) to obtain our final
predictions, we omit $z$ in the equations whenever possible to avoid clutter in the math.

\subsection{Predicting Halo Mass Distribution of Each Cluster}
\label{subsec:method_pmh}

To reveal the underlying dark matter halo population of each subsample of
clusters binned by $\msbcg$ and~(or) $\lambda$, we need to predict the halo mass
distribution of each cluster observed with $\msbcg$ and $\lambda$. Using Bayes' theorem,
we can write the PDF of halo mass as
\begin{equation}
    p(\mh \mid \msbcg, \lambda) = \frac{p(\msbcg, \lambda, \mh)}{p(\msbcg,
    \lambda)}=\frac{p(\msbcg, \lambda | \mh)\, p(\mh)}{p(\msbcg, \lambda)},
\end{equation}
using Equations~\ref{eqn:plntwod}, \ref{eqn:plntwodobs}, and $p(\mh)$.

Similarly, if we select clusters just by their richness, the PDF of halo mass is
simply
\begin{equation}
    p(\mh \mid \lambda) = \frac{p(\lambda | \mh)\, p(\mh) } {p(\lambda)},
\end{equation}
where
\begin{equation}
    p(\lambda) = \int_{M_h^{\mathrm{min}}}^{M_h^{\mathrm{max}}}  p(\lambda \mid
    \mh)\,p(\mh)\, \dd \mh.
\end{equation}
Therefore, the total number of $\lambda{\geq}20$ clusters within
$z_{\mathrm{min}}{=}0.17$ and $z_{\mathrm{max}}{=}0.30$ over a sky area $\Omega$
is
\begin{equation}
\langle N \rangle = \frac{n_0 \Omega}{4\pi}
    \int_{z_{\mathrm{min}}}^{z_{\mathrm{max}}}\frac{\dd V}{\dd
    z}\int_{\lambda_{\mathrm{min}}{=}20}p(\lambda | z)\,\dd \lambda.
    \label{eqn:N}
\end{equation}

More generically, we can compute the halo mass distribution of any set of $N$ clusters as
\begin{equation}
    p(\mh\mid \{\msbcg, \lambda\}_{i=1\cdots N}) =
    \frac{1}{N}\Sigma_{i=1}^{N}p(\mh|\{\msbcg, \lambda\}_i),
    \label{eqn:psum}
\end{equation}
and then we can predict the average halo mass of the same set of $N$ clusters as
\begin{equation}
    \langle \mh| \{\msbcg, \lambda\}_{i{=}1{\cdots}N}\rangle =
    \int \!\!p(\mh|\{\msbcg,
    \lambda\}_{i{=}1{\cdots}N}) \mh\,\dd \mh.
    \label{eqn:mavg}
\end{equation}
In particular, we predict the average halo mass of clusters with
$\lambda\in[\lambda_{\mathrm{min}}^j, \lambda_{\mathrm{max}}^j]$ by evaluating
Equations~\ref{eqn:psum} and \ref{eqn:mavg} over the $N_j$ clusters in each of
the four richness bins in Table~\ref{tab:wlmass}.

\section{Bayesian Inference: A Tale of Two Conformity Models}
\label{sec:infer}

\subsection{Model Degeneracy: Conformity vs. Anti-conformity}
\label{subsec:infer_degeneracy}

\begin{figure*}
\begin{center}
    \includegraphics[width=0.96\textwidth]{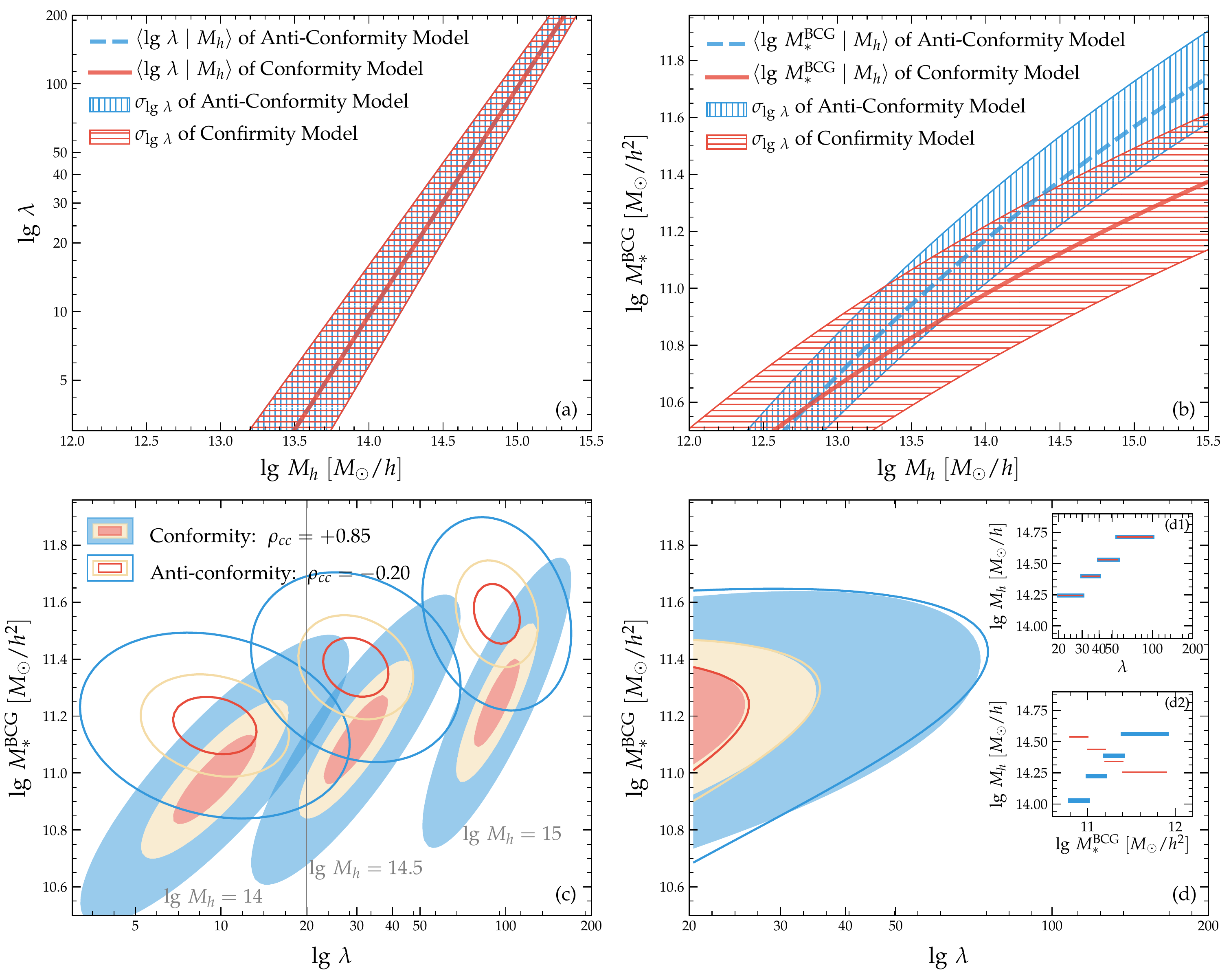} \caption{
	Pedagogical illustration of the degeneracy between
	conformity~($\rho_{\mathrm{cc}}>0$; red solid curves and filled
	contours) vs. anti-conformity~($\rho_{\mathrm{cc}}<0$; blue dashed
	curves and open contours) models of clusters. Panel (a): the mass--richness relations
    of the two models are exactly the same by design. Panel (b): the
    stellar-to-halo mass relations of the two models are different,
    with the conformity model having a shallower slope and larger scatter than
    the anti-conformity model. Panel (c): The three consecutive contours of each
    model indicate the 2D PDFs of clusters on the $\msbcg$ vs. $\lambda$ plane
    predicted at fixed log-halo masses of $14$, $14.5$, and $15$~(from left to right), respectively.
    The differences are not observable due to the lack of individual halo mass measurements.
    Panel (d): The 2D abundance of clusters on the observed $\msbcg$ vs. $\lambda$
    plane, predicted by the conformity~(filled) and anti-conformity~(open)
    models.  The inset panels (d1) and (d2) show the average halo mass of
    clusters in four bins of $\lambda$ and $\msbcg$, respectively. In each inset
    panel, thick blue and thin red lines indicate the predictions by the
    anti-conformity and conformity models, respectively. The two sets of model
    predictions are almost indistinguishable in their 2D abundances~(panel d)
    and halo mass in bins of $\lambda$~(panel d1), despite the large
    discrepancies shown in panels (b) and (c). This strong degeneracy can be
    potentially broken by measuring the halo mass of clusters in bins of
    $\msbcg$~(panel d2).  }
\label{fig:demo_dist2d}
\end{center}
\end{figure*}

Before moving on to the Bayesian inference of model parameters, we illustrate in
Figure~\ref{fig:demo_dist2d} that there exists a strong degeneracy in our
current model so that both conformity and anti-conformity models can describe
the 2D abundance of clusters and the weak lensing halo mass in bins of richness,
with exactly the same RHMR but different SHMRs.

In the top left panel of Figure~\ref{fig:demo_dist2d}, the red solid and blue
dashed lines are the mean RHMRs of the conformity and anti-conformity models,
respectively, with horizontally and vertically hatched bands of the same colours
indicating their corresponding scatters. The two RHMRs are exactly the same by
design, so that the two models will predict exactly the same average halo mass
for any cluster sample binned in richness~(as shown in Panel {\it d1}). The minimum richness
cut of $20$ is indicated by the gray horizontal line. In the top right panel of
Figure~\ref{fig:demo_dist2d}, we adopt the same plotting styles for the
conformity vs. anti-conformity models as in the top left panel, but show the
SHMRs instead. The SHMR of the conformity model~(red solid line with
horizontally hatched band) has a shallower slope but a larger scatter than that
of the anti-conformity model~(blue dashed line with vertically hatched band).
As a result, the two models will predict different average halo masses for clusters selected
by the BCG stellar mass~(as shown in Panel {\it d2}).

In the bottom left panel of Figure~\ref{fig:demo_dist2d}, filled and open
contours indicate the 2D PDFs of $\msbcg$ and $\lambda$ at
three fixed halo masses of $\lg\,\mh{=}14$, $14.5$, and $15$, of the
conformity~($\rho_{\mathrm{cc}}{=}0.85$) and anti-conformity~($\rho_{\mathrm{cc}}{=}-0.2$) models,
respectively. Each contour has three levels at $20\%$~(red), $50\%$~(beige), and
$90\%$~(blue) enclosed probabilities expanding outwards. Unsurprisingly, the two
models yield two drastically different $\plntwod$ at every mass. Yet, the 2D
abundance of clusters on the $\msbcg$ vs. $\lambda$ diagram~(bottom right panel)
predicted by the two models are strikingly similar --- the filled~(conformity)
and open~(anti-conformity) contours are mostly aligned and overlapping, leaving
little observational signature to distinguish the two models from 2D abundance alone. In the two inset
panels, we show the average halo mass of clusters in bins of $\lambda$~(panel {\it d1}) and
$\msbcg$~(panel {\it d2}), respectively. As expected, the halo masses are exactly the same
when binned by $\lambda$, as a result of the RHMRs being the same. Therefore, if
we constrain conformity using just the 2D abundance and halo mass in bins of
$\lambda$, there would be a strong degeneracy between the conformity and
anti-conformity models, as will be demonstrated later in
\S\ref{subsec:infer_degeneracy}.

However, the two sets of predicted halo mass in bins of $\msbcg$ are
significantly different. The conformity model predicts that the average halo
mass is a decreasing function of $\msbcg$, while the anti-conformity model
predicts a increasing trend with $\msbcg$. This discrepancy is likely related to the
``halo mass equality'' conundrum discovered in \citetalias{Zu2021}, showing that the
halo mass trend with $\msbcg$ could be highly non-trivial and depends critically
on the level of conformity within the cluster sample. More important, the strong
discrepancy shown in the panel {\it d2} of Figure~\ref{fig:demo_dist2d} implies that
the halo mass measurements for the low and
high-$\msbcg$ subsamples from \citetalias{Zu2021} could be the key in breaking
the degeneracy between the two types of conformity models, as will be shown
later in \S\ref{subsec:infer_resolve}. Note that the predictions of halo mass as a
function of stellar mass shown in panel {\it d2}~(and throughout this paper) are for
central galaxies of clusters with $\lambda{\geq}20$, therefore cannot be directly
compared with the measurements for central galaxies of {\it all} haloes~\citep{Mandelbaum2016, Zu2016}.

\subsection{Likelihood Model}
\label{subsec:infer_likelihood}

\begin{figure*}
\begin{center}
    \includegraphics[width=0.96\textwidth]{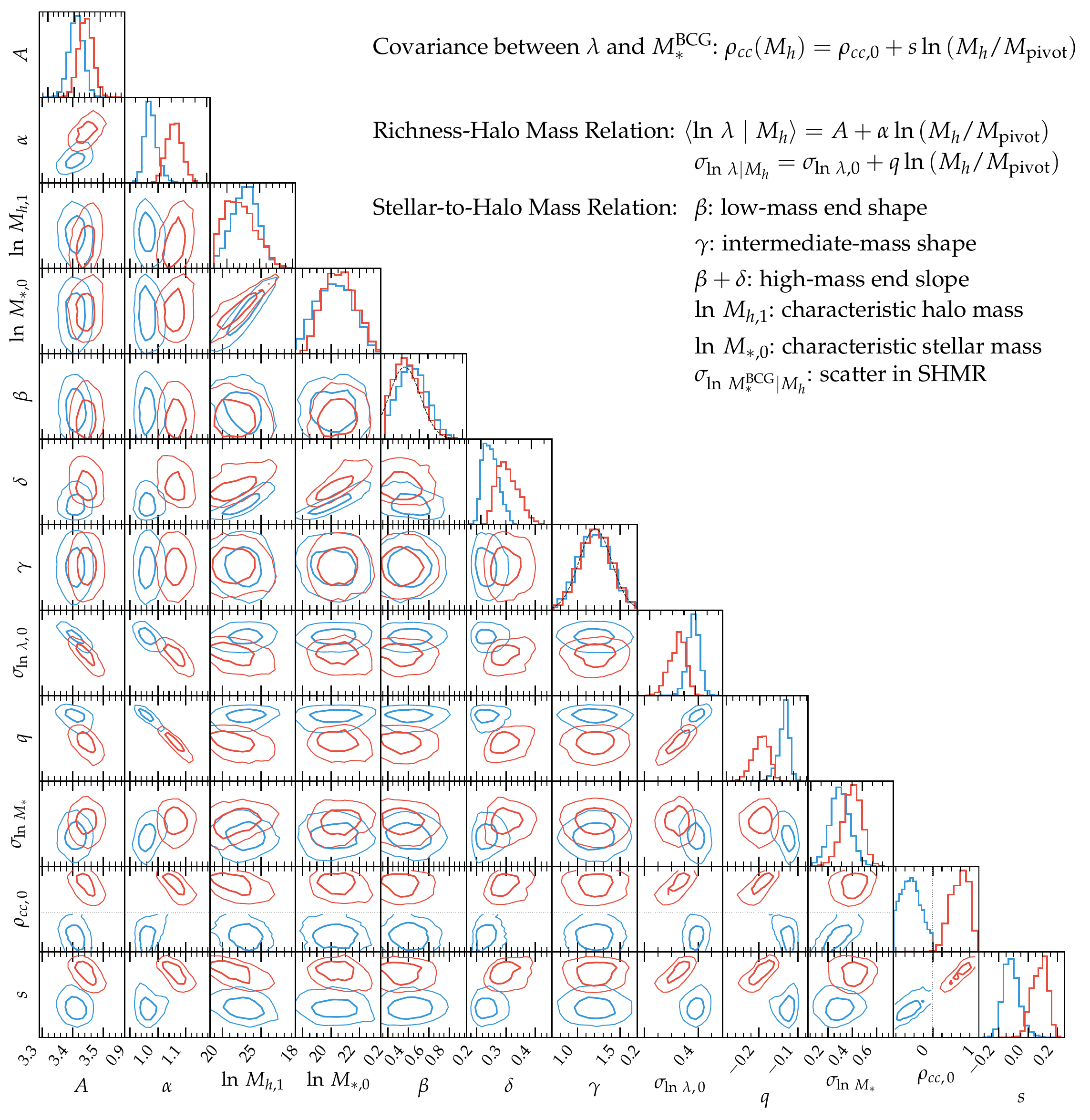} \caption{Parameter
    constraints of the conformity~(red) and anti-conformity~(blue) models.
    Diagonal panels show the 1D posterior distributions of each of the 12
    parameters, while the off-diagonal panels indicate the 2D confidence
    regions~(50 and 90 per cent from inside out) of the constraint on each of
    the parameter pairs. Gray dashed curves in the diagonal panels of $\beta$ and
    $\gamma$ are the Gaussian prior distributions. A short description of each
    parameter is given by the legend in the top right corner.}
\label{fig:glory}
\end{center}
\end{figure*}

To summarise our model parameters $\bm{\theta}$ from \S\ref{sec:method}, we have
in total 12 free parameters, including $\{A, \alpha, \siglamknot, q\}$ for
describing the RHMR, $M_{h, 1}, \{M_{*,0}, \beta, \delta, \gamma, \sigms\}$ for
describing the SHMR, and $\{\rho_{cc, 0}, s\}$ for describing the sign and level
of the BCG-satellite conformity.

Since $\beta$ and $\gamma$ describe the low-to-intermediate mass portion of the
SHMR, which is largely irrelevant to our constraint in the cluster mass regime,
we apply two Gaussian priors informed by the constraint from \citet{Zu2015}
using the galaxy clustering and galaxy-galaxy lensing measurements from SDSS:
$\beta{\sim}\mathcal{N}(0.33, 0.18^2)$ and $\gamma{\sim}\mathcal{N}(1.21,
0.19^2)$, respectively. For the rest of the parameters, we assume uniform priors
so that each parameter could vary freely within a range that is much larger than
potentially allowed by the data.

To recap our data vector from \S\ref{sec:data}, we have measured the total
number of observed clusters $N_{\mathrm{cls}}$, the BCG stellar mass and
satellite richness of individual clusters $\{\msbcg, \lambda\}_{i=1 \cdots
N_{\mathrm{cls}}}$, and the average halo mass of clusters binned in richness $\{
M_h \mid [\lambda_{\mathrm{min}}^j, \lambda_{\mathrm{max}}^j]\}_{j=1\cdots 4}$.
We will describe the likelihood model for each of the three components in turn
below.

We assume a Poisson likelihood model for $N_{\mathrm{cls}}$,
\begin{equation}
    \ln\,\mathcal{L}_{\mathrm{Pois}} = N_{\mathrm{cls}}\ln\langle N \rangle - \langle N
    \rangle - \ln \Gamma(N_{\mathrm{cls}}+1),
\end{equation}
where $\langle N \rangle$ is the expected total number of clusters predicted by
Equation~\ref{eqn:N} and $\Gamma$ is the Gamma function. For the 2D cluster
abundance, we simply multiply all the individual $\plntwodobs$ so that
\begin{equation}
    \ln\,\mathcal{L}_{\mathrm{Indi}} =  \Sigma_{i{=}1}^{N_{\mathrm{cls}}} \ln\,p(\ln
    M_{*, i}^{\texttt{BCG}}, \ln \lambda_i),
\end{equation}
where $p(\ln M_{*, i}^{\texttt{BCG}}, \ln \lambda_i)$ is the observing
probability of cluster $i$ derived from Equation~\ref{eqn:plntwodobs}. For the halo mass
in bins of richness, we assume a Gaussian likelihood model
\begin{equation}
    \ln\,\mathcal{L}_{\mathrm{Gaus}} = -\frac{1}{2}\ln\,|\textbfss{C}| - \frac{1}{2}
    (\bm{x} - \bm{\bar{x}})^T \textbfss{C}^{-1} (\bm{x} - \bm{\bar{x}}),
\end{equation}
where the $\bm{x}$ is the halo mass measurements for clusters in four richness
bins from Table~\ref{tab:wlmass}, $\bm{\bar{x}}$ is the predicted average halo
masses from Equation~\ref{eqn:mavg}, and $\textbfss{C}$ is the error matrix
associated with the halo mass measurements.

Finally, the full likelihood is the product of the three components
\begin{equation}
    p(\bm{y} \mid \bm{\theta}) = \mathcal{L}_{\mathrm{Pois}} \times \mathcal{L}_{\mathrm{Indi}} \times
    \mathcal{L}_{\mathrm{Gaus}},
\end{equation}
and the posterior probability is proportional to the product of the
likelihood and the prior probability $p(\bm{\theta})$
\begin{equation}
    p(\bm{\theta} \mid \bm{y}) \propto p(\bm{y} | \bm{\theta}) \, p(\bm{\theta}).
\end{equation}

\subsection{Parameter Constraint}
\label{subsec:infer_constraint}

\begin{figure*}
\begin{center}
    \includegraphics[width=0.96\textwidth]{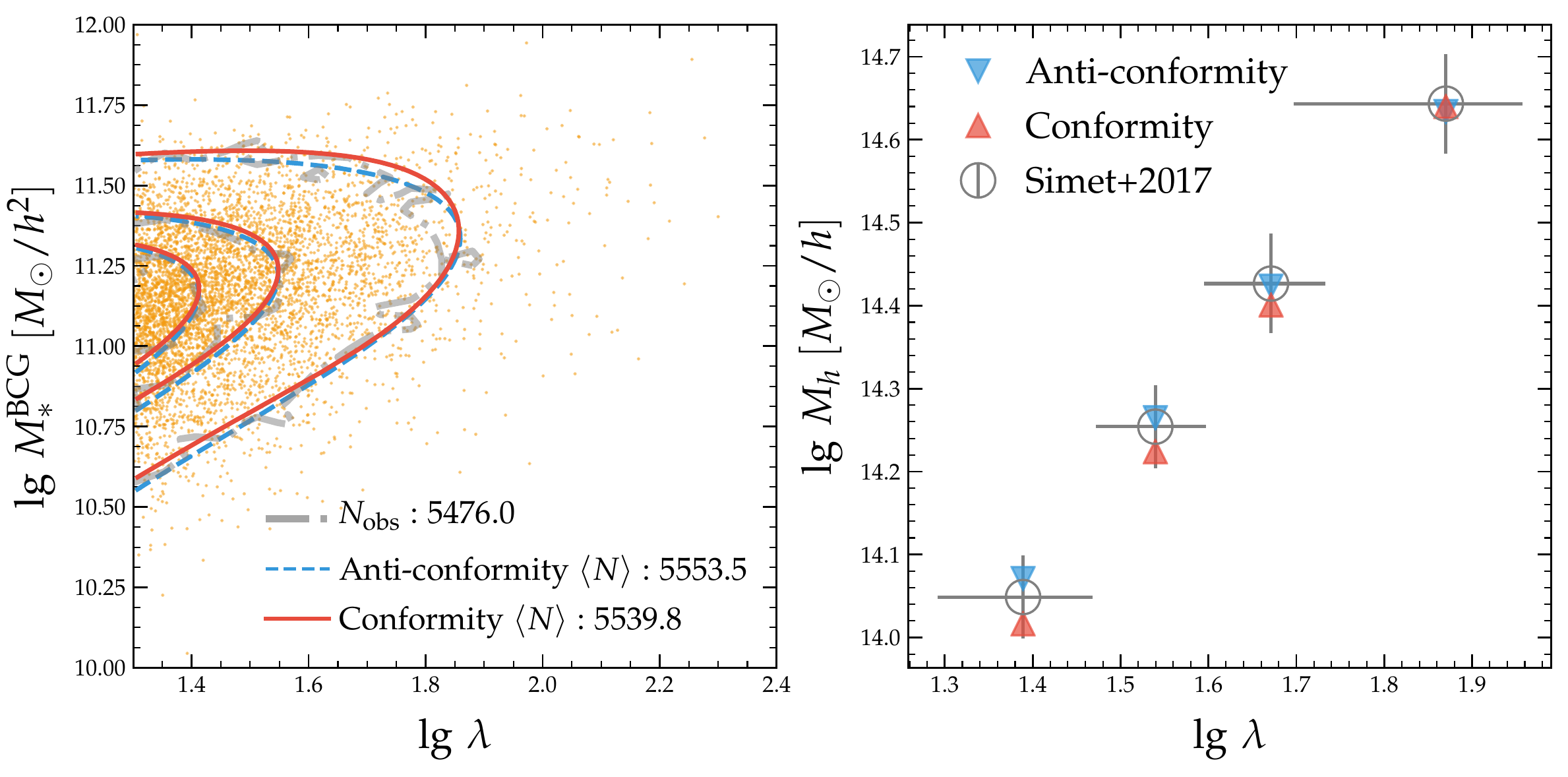}
    \caption{Comparison between the data and the predictions from the posterior
    mean parameters of the conformity~(red solid) and anti-conformity~(blue
    dashed) model constraints. {\it Left}: 2D PDFs of clusters on the $\msbcg$
    and $\lambda$ plane. Gray dot-dashed contours and yellow points indicate the
    observed PDFs and individual clusters, respectively. The total number of
    observed and predicted clusters are marked by the legend in the bottom
    right. {\it Right}: average log-halo mass in four bins of richness.
    Open circles with errorbars are the measurements from \citet{Simet2017}, and
    red and blue triangles indicate the posterior mean predictions from the
    conformity and anti-conformity models, respectively. Both models provide
    good fits to the data.}
\label{fig:data_vs_bestfit}
\end{center}
\end{figure*}

\begin{table}
\renewcommand*{\arraystretch}{1.25}
\centering \caption{Posterior constraints of the model parameters for the two models. The uncertainties are the $68\%$ confidence regions derived from the 1D posterior probability distributions.}
\begin{tabular}{ccc}
\hline
    Parameters              & Conformity Model           & Anti-conformity Model    \\
\hline
                 $A$        & $3.44_{-0.03}^{+0.03}$     & $3.41_{-0.03}^{+0.03}$   \\
            $\alpha$        & $1.06_{-0.03}^{+0.03}$     & $0.96_{-0.02}^{+0.03}$   \\
      $\ln\,M_{h,1}$        & $21.83_{-2.18}^{+2.60}$    & $22.82_{-2.10}^{+1.97}$  \\
      $\ln\,M_{*,0}$        & $20.57_{-1.13}^{+1.20}$    & $20.23_{-1.36}^{+1.31}$  \\
             $\beta$        & $0.35_{-0.15}^{+0.18}$     & $0.42_{-0.17}^{+0.18}$   \\
            $\delta$        & $0.30_{-0.04}^{+0.05}$     & $0.24_{-0.03}^{+0.03}$   \\
            $\gamma$        & $1.21_{-0.19}^{+0.19}$     & $1.20_{-0.19}^{+0.20}$   \\
    $\sigma_{\ln\,\lambda,\,0}$ & $0.32_{-0.04}^{+0.03}$ & $0.39_{-0.03}^{+0.03}$   \\
                 $q$        & $-0.20_{-0.03}^{+0.02}$    & $-0.13_{-0.02}^{+0.01}$  \\
 $\sigma_{\ln\,M_*}$        & $0.40_{-0.10}^{+0.09}$     & $0.28_{-0.09}^{+0.10}$   \\
     $\rho_{cc,\,0}$        & $0.60_{-0.23}^{+0.17}$     & $-0.48_{-0.22}^{+0.24}$  \\
                 $s$        & $0.08_{-0.06}^{+0.05}$     & $-0.12_{-0.05}^{+0.06}$  \\
\hline
\end{tabular}
\label{tab:constraint}
\end{table}

Equipped with the full likelihood model, we now set out to infer the joint
posterior distribution of the 12 model parameters. We first perform the analysis
by allowing $\rho_{cc,0}$ to vary freely between $-1$ and $1$, yielding a
dominant solution that prefers a positive $\rho_{cc,0}$~(i.e., conformity), as
well as a secondary solution with $\rho_{cc,0}{<}0$~(i.e., anti-conformity).
This model degeneracy is expected from our simple experiment in
\S\ref{subsec:infer_degeneracy}. To thoroughly explore the two solutions
separately, we then repeat our inference twice by first limiting
$\rho_{cc,0}{\in}[0, 1)$ and then $\rho_{cc,0}{\in}(-1, 0]$ when sampling the
posterior distributions, yielding a best--fitting conformity and an
anti-conformity model, respectively.  Since we will be distinguishing the two
degenerate solutions using the weak lensing of clusters binned by $\msbcg$
in~\S\ref{subsec:infer_resolve}, we present the two solutions in parallel below without comparing their relative statistical
significance.

For each inference, we employ the affine invariant Markov Chain Monte
Carlo~(MCMC) ensemble sampler \texttt{emcee}~\citep{Foreman-Mackey2013}.  We run
the MCMC sampler for $2,000,000$ steps for each analysis to ensure its
convergence, and derive the posterior constraints after a burn-in period of
$500,000$ steps. The median values and the 68 per cent confidence limits of the
1D posterior constraints are listed in Table~\ref{tab:constraint}.

Figure~\ref{fig:glory} compares the two separate parameter constraints derived
for the conformity~(red) and anti-conformity~(blue) models. For each model, the
histograms in the diagonal panels show the 1D marginalised posterior
distributions of each of the 12 parameters,
and the contours in the off-diagonal panels are the $50\%$ and $90\%$ confidence regions
for each of the parameter pairs.  The gray solid line running through the
$\rho_{cc,0}$-related panels divides the conformity vs. anti-conformity models
at $\rho_{cc,0}{=}0$.  In the top right corners, we provide a brief description
of the functionality of each model parameter within each of the three components
of our $\plntwod$ model~(i.e., RHMR, SHMR, and BCG-satellite conformity).

Figure~\ref{fig:data_vs_bestfit} compares the predictions from the
conformity~(red) and anti-conformity~(blue) posterior mean models to the data.
The orange dots in the left panel represent the observed cluster distribution on
the $\msbcg$ vs. $\lambda$ diagram, with the three thick gray dashed contour
lines enclosing $20$, $50$, and $90$ percentiles of the cluster sample~(from the
inside out), respectively. Red solid and blue dashed contour lines indicate the
same three levels in percentiles predicted by the conformity and anti-conformity
posterior mean models. Both models provide adequate descriptions of the
underlying 2D distributions of clusters on the $\msbcg$ vs. $\lambda$ diagram.
The right panel of Figure~\ref{fig:data_vs_bestfit} compares the weak
lensing-measured halo masses in four bins of richness~(open circles with
errorbars) to those predicted by the conformity~(red triangles) and
anti-conformity~(blue inverted triangles) posterior mean models. Both model
predictions are in good agreement with the weak lensing mass measurements, with
the (anti-)conformity model predictions slightly lower~(higher) than the
observations at the low richness end. Overall, Figure~\ref{fig:data_vs_bestfit}
confirms our expectation from Figure~\ref{fig:demo_dist2d} that there exists a model
degeneracy that cannot be overcome by the combination of 2D cluster abundance and
halo mass measurements in bins of richness.

\begin{figure*}
\begin{center}
    \includegraphics[width=0.96\textwidth]{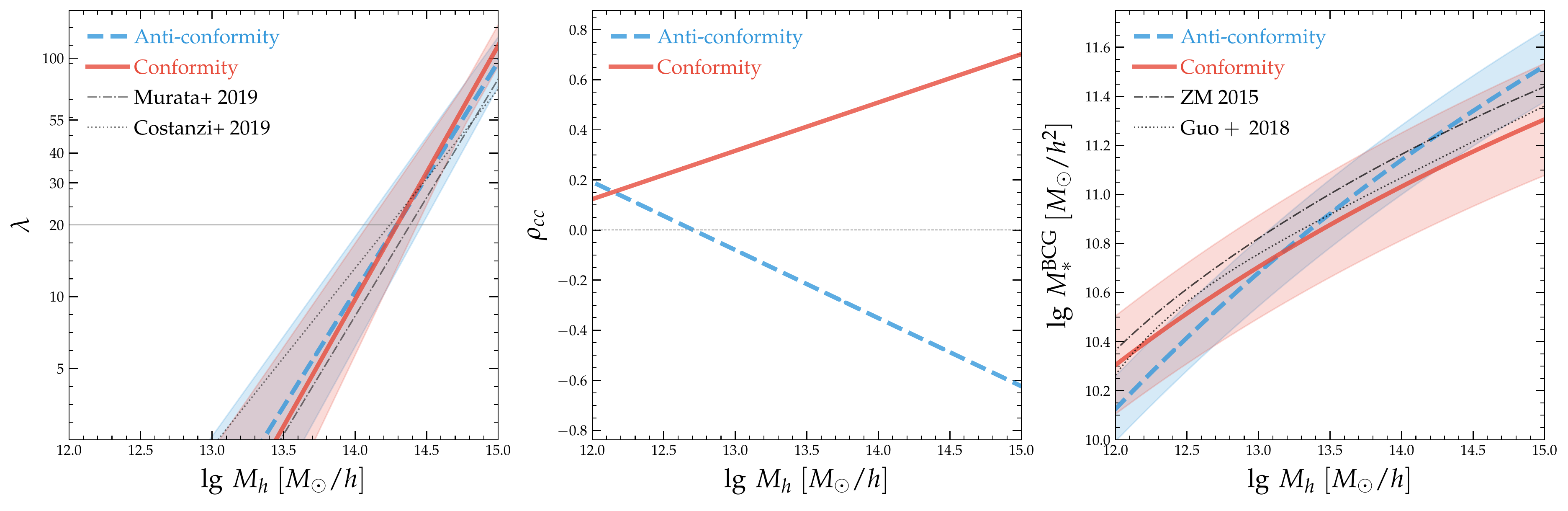}
    \caption{Mass--richness relation~(left), correlation coefficient
    variation~(middle), and stellar-to-halo mass relation~(right), predicted by
    the conformity~(red solid) and anti-conformity~(blue dashed) posterior mean
    models.  Dotted and dot-dashed lines in the left panel indicate the
    best--fitting models from \citet{Costanzi2019} and \citet{Murata2018},
    respectively. In the right panel, dotted and dot-dashed curves are the
    best--fitting models from \citet{Guo2018} and \citet{Zu2015}, respectively.
    Shaded bands in the left and right panels indicate the $1{-}\sigma$
    logarithmic scatter about the median scaling relations.}
\label{fig:scaling_relations}
\end{center}
\end{figure*}

Figure~\ref{fig:scaling_relations} provides a more visually-appealing way of
comparing the two model constraints. Instead of showing the posterior
distributions of the 12 individual parameters, we examine the behaviors of the
best-fitting RHMRs~(left), BCG-satellite conformities~(middle), and
SHMRs~(right), respectively. In the left panel, we also show the mean RHMRs derived
by \citet{Murata2018}~(gray dot dashed) and \citet{Costanzi2019}~(gray dotted)
for the SDSS redMaPPer clusters. Unsurprisingly, the RHMRs predicted by the
conformity~(red solid) and anti-conformity~(blue dashed) models are reasonably
similar at $\lambda{\geq}20$, because they are primarily constrained by the
observed abundance and average halo mass of clusters binned in richness, which are independent of $\rho_{\mathrm{cc}}(\mh)$. The shaded band about each
mean RHMR indicates the dependence of scatter on halo mass.
Compared to the \citet{Murata2018}
result, our RHMRs have a higher amplitude but the same slope, likely due to a
slight shift in the weak lensing halo mass calibration compared to ours --- they
used the shear catalogue from the HSC survey and a cluster sample with
$z\in[0.1, 0.33]$. The RHMR derived by \citet{Costanzi2019} has a much shallower
slope than the other three, because they also varied cosmology while inferring
the RHMR.

In the right panel of Figure~\ref{fig:scaling_relations}, we compare the SHMRs
inferred from the conformity~(red solid) and anti-conformity~(blue dashed)
models, as well as the results from \citet{Zu2015}~(gray dot-dashed) using the
galaxy clustering and galaxy--galaxy lensing at $z{\sim}0.1$ and from
\cite{Guo2018}~(gray dotted) using the LOWZ galaxy clustering at the same
redshift of our sample. Unlike the RHMRs, our two inferred SHMRs are significantly
different, with the conformity SHMR showing a shallower slope but a larger
scatter than the anti-conformity one. Additionally, the result from
\citet{Guo2018} is consistent with the conformity constraint, while the
\citet{Zu2015} curve has a similar slope but a higher amplitude compared to the
conformity prediction, probably due to some redshift evolution of the SHMR from
$z{\sim}0.25$ to $0.1$. Both the \citet{Guo2018} and \citet{Zu2015} constraints
are strongly inconsistent with the prediction by the anti-conformity model.

Finally, the middle panel of Figure~\ref{fig:scaling_relations} shows the
correlation coefficients as functions of halo mass $\rho_{\mathrm{cc}}(\mh)$,
predicted by the conformity~(red solid) and anti-conformity~(blue dashed)
models, respectively. Interestingly, both predictions favor a weak correlation
between the BCG stellar mass and satellite richness at the low mass end~(i.e.,
below $10^{13}\,\hmsol$), but bifurcate into strong positive and negative
correlations at the high mass end~(i.e., above a few times $10^{14}\,\hmsol$).
The two different scenarios point to drastically different paths of galaxy
formation within massive clusters --- the conformity model implies a {\it
correlated} growth between the BCG and satellite galaxies, while the anti-conformity
model favors a {\it compensated} growth between the two galaxy populations.
Therefore, it is vital to observationally distinguish the two scenarios for a
better understanding of the underlying physics behind cluster galaxy formation.

\subsection{Resolving the ``Halo Mass Equality'' Conundrum}
\label{subsec:infer_resolve}

\begin{figure*}
\begin{center}
    \includegraphics[width=0.96\textwidth]{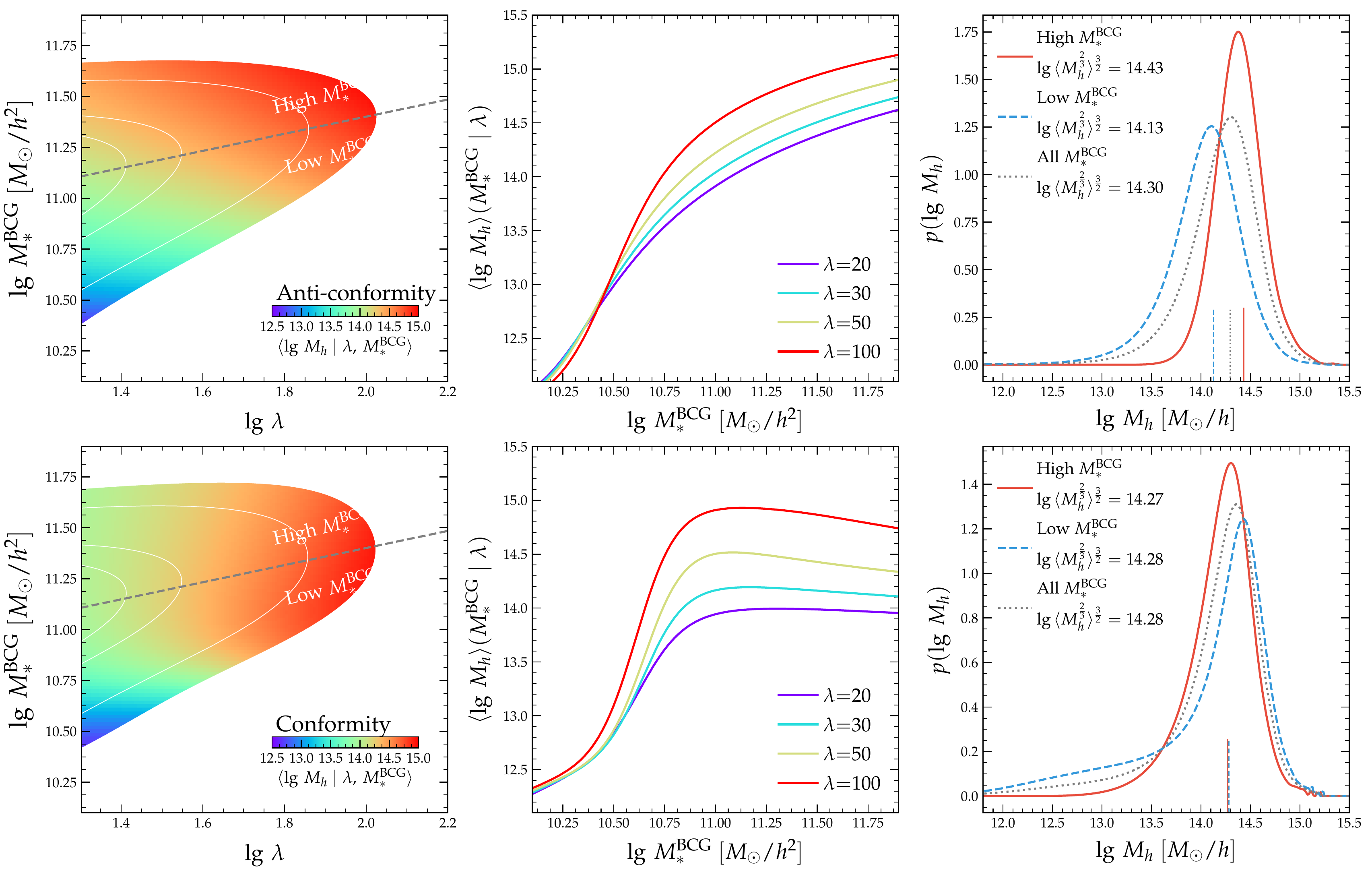}
    \caption{Predictions of halo mass distributions by the anti-conformity~(top
    row) and conformity~(bottom row) posterior mean models. In each row, the
    left panel shows the map of average log-halo mass on the $\msbcg$ vs.
    $\lambda$ plane, colour-coded by the horizontal colourbar in the bottom right corner.
    Gray dashed line indicates the median log-$\msbcg$ vs. $\lambda$ relation,
    which divides the clusters into two subsamples of high and low $\msbcg$ with the same distribution of $\lambda$. The middle panel shows the
    variation of log-$\mh$ as a function of $\msbcg$, at four fixed values of
    $\lambda$ of 20~(purple), 30~(cyan), 50~(yellow), and 100~(red),
    respectively. The right panel compares the PDFs of log-$\mh$ of the
    high-$\msbcg$~(red solid) and low-$\msbcg$~(blue dashed) subsamples, as
    defined in the left panel, with the PDF of all the clusters shown by the
    gray dotted curve. The predicted average weak lensing halo mass of each
    cluster (sub)sample is marked by the bottom vertical tick of the respective
    line color/style and shown in the legend. The two models predict significantly
    different halo masses for the high and low-$\msbcg$ subsamples, despite
    reproducing the similar
    observables in Figure~\ref{fig:data_vs_bestfit}.}
\label{fig:mstar_split}
\end{center}
\end{figure*}

As mentioned in the introduction, we are hopeful that the existence of a strong
(anti-)conformity between the BCG stellar mass and satellite richness could
potentially reconcile the ``halo mass equality'' conundrum  --- that is, when
split into two halves by the median $\msbcg$ at fixed $\lambda$, the two cluster
subsamples have almost the same average halo mass, despite having a $0.34$ dex
discrepancy in their average $\msbcg$.  We refer readers to~\citetalias{Zu2021}
for details on the subsample definition~(see also Figure~\ref{fig:mstar_split})
and the original ``halo mass equality'' conundrum. We now explore whether one of
the two posterior mean models we inferred in~\S\ref{subsec:infer_constraint} is
consistent with such ``halo mass equality'' phenomenon.

Figure~\ref{fig:mstar_split} illustrates the differences between the
anti-conformity~(top row) and conformity~(bottom row) models in decomposing
the underlying halo mass distribution of the high and low-$\msbcg$
subsamples.  In each row, the left panel shows the variation of the average
halo mass across the $\msbcg$ vs. $\lambda$ plane, with the logarithmic
mass indicated by the colourbar in the bottom right. The gray dashed line
represents the median $\msbcg$--$\lambda$ relation that splits the clusters
into high and low-$\msbcg$ subsamples in~\citetalias{Zu2021}.
Interestingly, the high-$\msbcg$ clusters have on average higher halo
masses than the low-$\msbcg$ systems in the anti-conformity scenario~(top
left), while the trend of average halo mass with $\msbcg$ is less clear in
the conformity model~(bottom left). We further clarify the halo mass trend
with $\msbcg$ in the middle panels, by showing the average log-halo mass as
functions of $\msbcg$ at four different richnesses of $20$~(purple),
$30$~(green), $50$~(yellow), and $100$~(red), respectively.  Clearly, all
the four curves are monotonic with $\msbcg$ in the anti-conformity model,
but exhibit a plateau above $\msbcg{\sim}10^{11}\hhmsol$ in the conformity
model. Note that the plateau is a unique feature predicted by the
conformity, and cannot be mimicked by systematic effects like the
mis-centring, which primarily affects the low-$\msbcg$ systems.

The right panels of Figure~\ref{fig:mstar_split} provide the key to potentially
resolving the ``halo mass equality'' conundrum. In each panel, we show the
underlying halo mass distributions for all~(gray dotted), low-$\msbcg$~(blue
dashed), and high-$\msbcg$~(red solid) clusters, respectively. Additionally, we
indicate the average weak lensing halo mass~\citep[$\langle M_h^{2/3}
\rangle^{3/2}$; see][]{Mandelbaum2016} of each of the three distributions using
a short vertical line of the same colour at the bottom. Unsurprisingly, the two
subsamples are predicted to have a ${\sim}0.3$ dex discrepancy in their average weak lensing
halo mass in the anti-conformity model~(top right), due to the monotonic trend
of halo mass with $\msbcg$ across the entire richness range.

However, in the bottom right panel of Figure~\ref{fig:mstar_split}, the
conformity model predictions exhibit exactly the same ``halo equality'' as
discovered in~\citetalias{Zu2021} --- the two subsamples of clusters have
almost the same weak lensing halo mass, despite the significant difference
in the shape of their halo mass distributions.  In particular, the
conformity model predicts a stronger low-$\mh$ tail and a more massive peak
for the halo mass distribution of the low-$\msbcg$ subsample than the
high-$\msbcg$ one. More important, the low-$\mh$ tail and the high-$\mh$
peak somehow conspire to produce an average weak lensing halo mass that is
very similar to that of the high-$\msbcg$ clusters, thereby resolving the
``halo mass equality'' conundrum of~\citetalias{Zu2021}.

The average halo mass estimated by \citetalias{Zu2021} for the low and
high-$\msbcg$ subsamples is $\lg\mh{=}14.24{\pm}0.02$, roughly $10\%$ lower
than predicted by the best--fitting conformity model~($14.28$). However, as
mentioned in~\S\ref{subsec:data_clusterwl}, the estimated uncertainty of
halo mass~(${\pm}0.02$ dex) in \citetalias{Zu2021} does not include many of
the systematic uncertainties that were included by \citet{Simet2017},
therefore should be considered a lower limit. If we assume the typical mass
error of $0.05$ dex from \citet{Simet2017}, e.g., by adding an extra $0.03$
dex of fully-correlated systematic error, the two sets of halo mass
estimates would be consistent within $1\sigma$. In the future, we can
further tighten the constraints on conformity by applying a uniform halo
mass measurement method to clusters binned by $\msbcg$ and $\lambda$.

Before moving on to the second half of the paper, we summarise our key
results so far as follows.
\begin{itemize}
    \item We have inferred the best-fitting models under different
    assumptions of conformity vs. anti-conformity between the BCG stellar
    mass and satellite richness, using the combination of cluster abundance
    and weak lensing mass of clusters binned in $\lambda$ as constraints.
    \item Both best-fitting conformity and anti-conformity models provide
    good descriptions of the data, but they predict significantly different
    average halo masses for clusters binned in $\msbcg$.
    \item By comparing to the weak lensing halo mass measurements of the
    low and high-$\msbcg$ clusters, we demonstrated that while the
    anti-conformity model is strongly disfavored by the data, the
    best--fitting conformity model predicts the same average halo mass for
    the two cluster subsamples, thereby resolving the ``halo mass
    equality'' conundrum discovered by
    \citetalias{Zu2021}~(Figure~\ref{fig:mstar_split}).
\end{itemize}

\section{Modelling Cluster Weak Lensing with Conformity and Assembly Bias}
\label{sec:dsfull}

Apart from the ``halo mass equality'' conundrum that we focused on in the
first part of this paper, \citetalias{Zu2021} also
discovered that the low-$\msbcg$ clusters on average exhibit a $20\%$ lower
concentration~($5.87$ vs.  $6.95$) and a ${\sim}10\%$ higher large scale
bias than their low-$\msbcg$ counterparts. \citetalias{Zu2021} suggested
that the bias discrepancy could be an evidence of cluster assembly
bias~\citep{Zu2017}.  However, while the concentration measurements from
the small--scale weak lensing profiles are robust~(modulo the degeneracy
with the cluster mis-centering effect), the modelling of large--scale
biases in~\citetalias{Zu2021} is lacking, due to the omission of the halo
assembly bias effect that governs the concentration--bias relation of
clusters at fixed $\mh$.

Therefore, in the second part of this paper we will implement halo assembly
bias in our posterior mean conformity model inferred from
\S\ref{subsec:infer_constraint}, in order to provide a more accurate model
for the weak lensing profiles $\ds$ which can then be compared with the
measurements for the low and high-$\msbcg$ subsamples from
\citetalias{Zu2021}.

To avoid distracting the impatient, we will directly present the $\ds$
predictions from our best-fitting analytic models that include the halo
assembly bias and (anti-)conformity in this section. For those who are
interested in the modelling details, we describe the calibration and
prescription of halo assembly bias in Appendix \S\ref{sec:dsfull_cab}, and
the analytic model of weak lensing profiles in Appendix
\S\ref{sec:dsfull_joint}.

\subsection{Fitting to Weak Lensing of High and Low-\texorpdfstring{$\bm{\msbcg}$}{Mstar} Clusters}
\label{subsec:dsfull_fit}

\begin{figure*}
\begin{center}
    \includegraphics[width=0.96\textwidth]{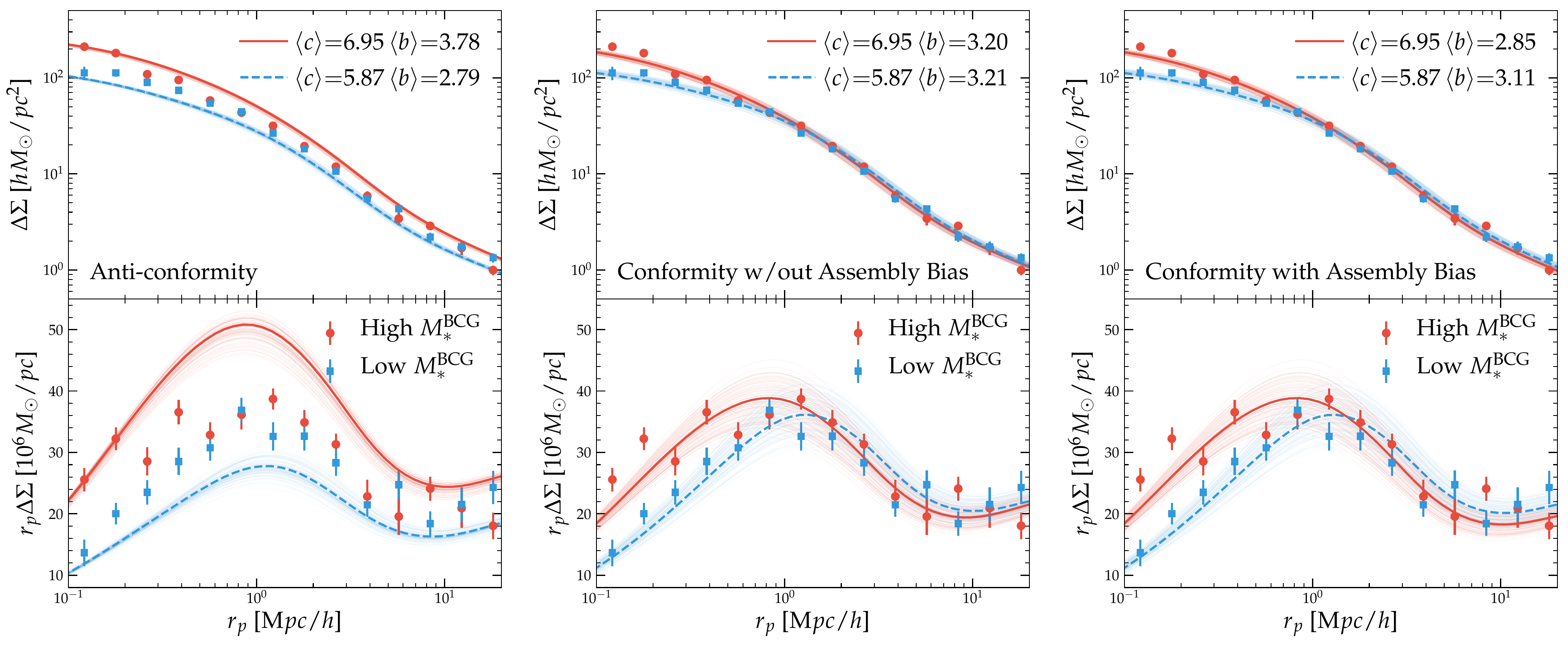}
    \caption{Comparison between the surface density contrast profile
    $\Delta\Sigma$ measured by~\citetalias{Zu2021} from weak lensing and
    that predicted by three different models: anti-conformity~(left
    column), conformity without assembly bias~(middle column), and
    conformity with assembly bias~(right column). In each column, the top
    and bottom panels show the same information except that the y-axis of
    the bottom panel is $r_p\Delta\Sigma$. In each panel, red circles and
    blue squares with errorbars are the weak lensing measurements for the
    high and low-$\msbcg$ subsamples, respectively. Red solid and blue
    dashed thick curves are the predictions from the respective posterior
    mean model, while the bundle of thin curves around each thick curve is
    predicted from 100 random steps along the respective MCMC chain from
    Figure~\ref{fig:glory}.  The values of average concentration and bias
    adopted by each model are listed in the top right corner of each
    column.  For the $\Delta\Sigma$ prediction, we adjust the average halo
    concentrations to be the best--fitting values inferred
    from~\citetalias{Zu2021} on small scales, but predict the large-scale
    bias using different prescriptions~(see text for details).
    }
\label{fig:ds_split}
\end{center}
\end{figure*}

Figure~\ref{fig:ds_split} compares the stacked weak lensing measurements to
those predicted by the posterior mean anti-conformity model~(left column),
conformity model without assembly bias~(middle column), and conformity
model with assembly bias~(right column), respectively. The top and bottom
rows are the same except for the labels of the y-axes~($\ds$ vs. $r_p\ds$).
In each panel, red circles and blue squares are the weak lensing
measurements of the high and low-$\msbcg$ subsamples~(same as those shown
in the Figure 5 of~\citetalias{Zu2021}), respectively, while red solid and
blue dashed curves are the respective model predictions. The values of
average concentration $\avg{c}$ and average bias $\avg{b}$ adopted by each
model are indicated in the top right of each top panel. Unsurprisingly, the
predictions by the anti-conformity model in the left panels fail to
describe the $\ds$ measurements on all scales, due to the factor of two
difference between the two predicted
average halo masses and the ${\sim}30\%$ discrepancy between the two predicted biases.

In the middle panels of Figure~\ref{fig:ds_split}, the conformity-only~(i.e.,
without assembly bias) model provides good description of the small-scale $\ds$
measurements, echoing the finding in Figure~\ref{fig:mstar_split} that the average
weak lensing halo masses of the two subsamples are similar. On large scales, the
two predicted $\ds$ profiles converge to the same amplitudes, because the
similar average halo masses produce similar biases in the absence of
assembly bias. In the right panels of Figure~\ref{fig:ds_split}, the small-scale
behavior of the $\ds$ profiles predicted by the conformity+AB~(i.e., with
assembly bias) model are the same as in the middle panels, but on large scales
the two predicted curves differ by about $10\%$ --- the high-$\msbcg$ clusters are
more concentrated, producing a lower bias than the low-$\msbcg$ systems due to
the cluster assembly bias effect.

\begin{figure*}
\begin{center}
    \includegraphics[width=0.96\textwidth]{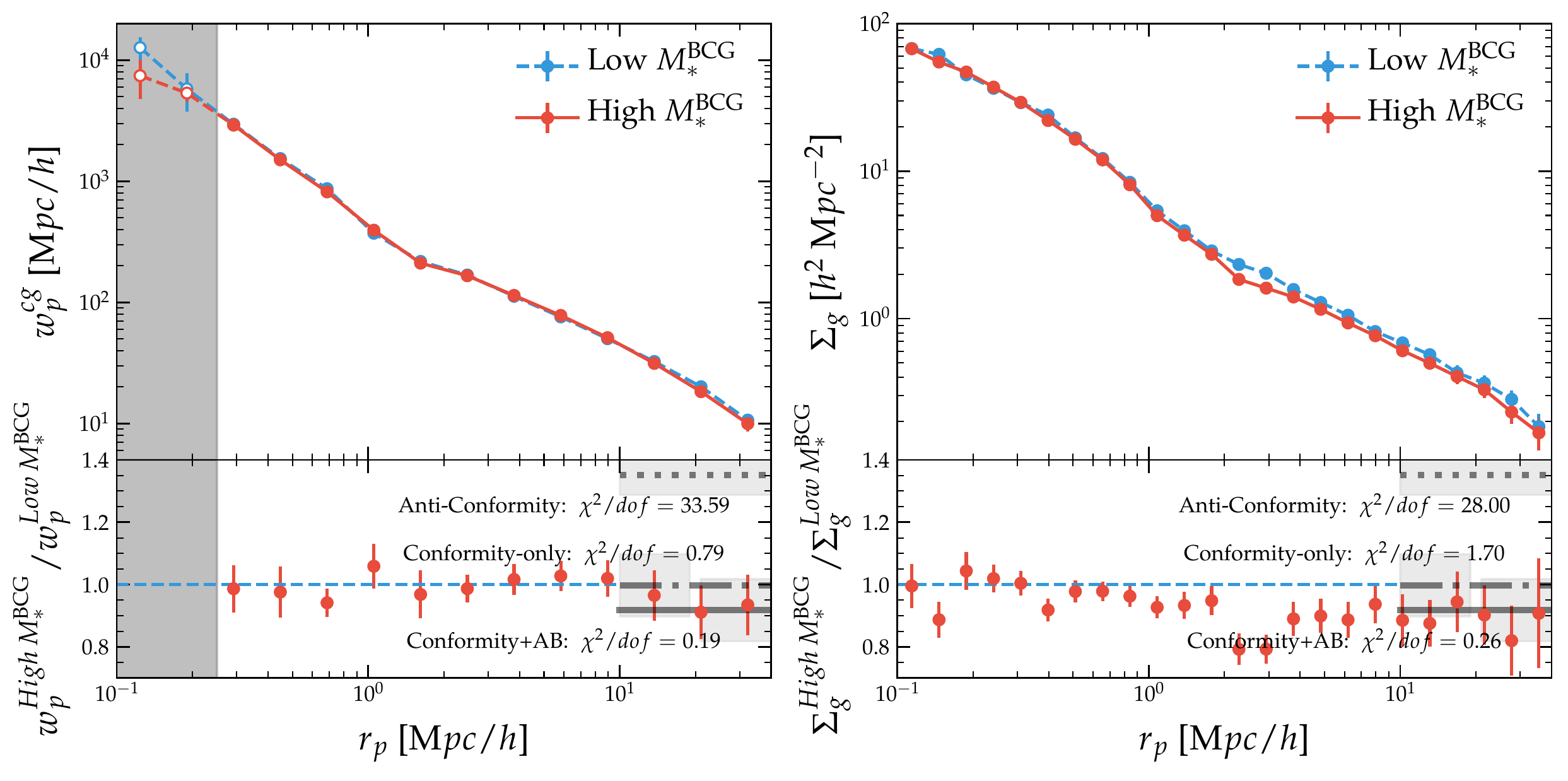}
    \caption{{\it Left panel}: Comparison of the projected correlation
    functions $w_p^{cg}$ of the low~(blue) and high~(red) $\msbcg$
    subsamples with the BOSS LOWZ spectroscopic galaxy sample in the upper
    sub-panel, with the ratio of the two shown in the bottom sub-panel. The
    gray shaded region on the left indicates the distance scales that are
    affected by the fibre collision in BOSS. {\it Right panel}: Similar to
    the left panel, but for the excess galaxy surface number density
    profile $\Sigma_g$ calculated from the SDSS DR8 imaging.  In each
    sub-panel, the bias ratios predicted by the anti-conformity,
    conformity-only~(i.e., without assembly bias), and conformity+AB~(i.e.,
    with assembly bias) models are indicated by the gray horizontal dotted,
    dot-dashed, and solid lines on scales above $10\hmpc$, respectively.
    The gray shaded band around each horizontal line indicates the
    1$\sigma$ uncertainty predicted by the constraints from
    Figure~\ref{fig:glory}.
    }
\label{fig:wpsignals_marked}
\end{center}
\end{figure*}

Due to the relatively large errorbars of $\ds$ on large scales, it is
difficult to ascertain whether the conformity+AB model is superior to the
conformity-only model.  Therefore, we further examine the large-scale
behaviors of the three models in Figure~\ref{fig:wpsignals_marked}, where
we show the projected cross--correlation functions between clusters and
LOWZ galaxies~(left) and the cluster galaxy number density profiles
measured from the cross--correlations with photometric galaxies~(right).
Although we cannot directly measure the cluster biases directly from the
cross-correlations with galaxies, which also depend on the bias of the
galaxies\citep{Xu2021}, we can distinguish the three models by examining
the ratio between the cross-correlations of the high and low-$\msbcg$ with
galaxies, which is a direct measure of $b_+/b_-$ independent of galaxy
bias.

In each panel of Figure~\ref{fig:wpsignals_marked}, red and blue circles with
errorbars indicate the measurements for the high and low-$\msbcg$ subsamples,
respectively. In the bottom sub-panel, red circles are the ratio between the
measurements of the two subsamples. The gray shaded region in the left panel
indicates the projected distances that are affected by the fibre collision in
BOSS. Figure~\ref{fig:wpsignals_marked} is the same as the Figure 6
of~\citetalias{Zu2021}, except that we mark the large-scale ratios predicted by
the anti-conformity~(dotted horizontal line), conformity-only~(dot-dashed), and
conformity+AB~(solid) models in the bottom sub-panels. Clearly, the
anti-conformity prediction is ruled out by the data. The conformity-only
prediction without assembly bias is also disfavored by the
observations, which exhibit a $10\%$ bias discrepancy between the two subsamples.
Meanwhile, the direction and amplitude of this bias discrepancy is in good agreement with the prediction by the
conformity model with assembly bias. This is very reassuring --- the combination
of BCG-satellite conformity and cluster assembly bias not only predicts the
correct weak lensing masses of clusters selected by $\msbcg$,
therefore resolving the intriguing conundrum discovered in~\citetalias{Zu2021},
but also accurately reproduces the large-scale bias inversion with $\msbcg$
using the cluster assembly bias model directly predicted by the $\lcdm$ simulations.

\subsection{Exploring Projection Effects}
\label{subsec:dsfull_proj}

The projection effects in photometric cluster detection could induce systematic
errors in the cluster observables that could sometimes masquerade as physical
phenomena~\citep{Zu2017, Busch2017}. As mentioned in the introduction,
\citet{To2020} discussed the possibility of projection effects to induce a
positive correlation between BCG luminosity and richness, by enhancing the
estimated richness in the dense environments that potentially host older and
more luminous BCGs at fixed halo mass. If the project effects are indeed the
culprit, we should expect some correlation between the BCG stellar mass and the
level of cluster membership contamination due to projection effects.

To investigate whether the strong conformity and assembly bias signals are
partly induced by projection effects, we adopt the average membership distance
$R_{\mathrm{mem}}$ as our measure of the projection effect in each cluster, defined as
\begin{equation}
    R_{\mathrm{mem}}= \frac{\sum_i (p^{i}_{m} \, R_i)}{\sum_i p^{i}_{m}},
    \label{eqn:defrmiya}
\end{equation}
where $p_m^i$ and $R_i$ are the membership probability and the projected
distance from the BCG of the $i$-th member galaxy candidate in that cluster,
respectively. \citet{Miyatake2016} initially used $R_{\mathrm{mem}}$ as a proxy
for halo concentration but found an extremely high signal of cluster assembly
bias that is inconsistent with the $\lcdm$ simulations. \citet{Zu2017} later
demonstrated that $R_{\mathrm{mem}}$ is strongly correlated with the fraction of
spurious member galaxies in each cluster, causing the inconsistency between the
\citet{Miyatake2016} measurement and $\lcdm$. Therefore, we expect
$R_{\mathrm{mem}}$ to be a good indicator of the level of membership
contamination in individual clusters.

\begin{figure*}
\begin{center}
    \includegraphics[width=0.96\textwidth]{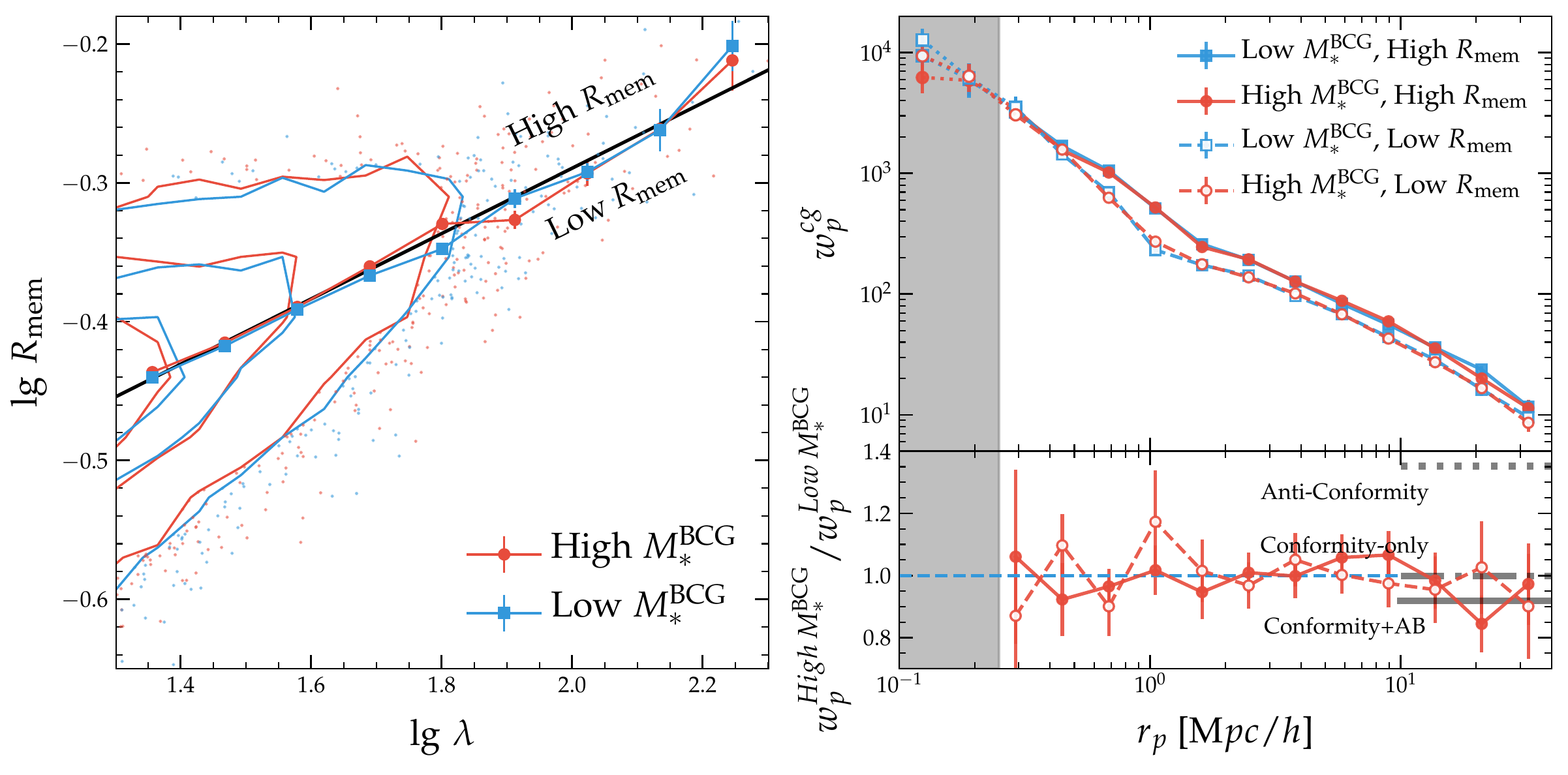}
    \caption{Impact of projection effects on the detection of conformity
    and assembly bias, using the average member galaxy distance
    $R_{\mathrm{mem}}$ as a proxy for the strength of projection effects.
    {\it Left}: Distributions of low~(blue) and high~(red) $\msbcg$
    clusters on the $R_{\mathrm{mem}}$ vs. $\lambda$
    plane. Each set of contour lines indicate the $20\%$, $50\%$, and $90\%$ enclosed
    regions from the inside out. Red circles and blue squares show the median
    $R_{\mathrm{mem}}$ at fixed $\lambda$ of the low and high-$\msbcg$
    subsamples, with the errorbars indicating the uncertainties on the median.
    The solid black line is a power-law fit to the two median relations,
    dividing the clusters into low and high-$R_{\mathrm{mem}}$ populations. {\it
    Right}: Similar to the left panel of Figure~\ref{fig:wpsignals_marked}, but
    for subsamples further split by $R_{\mathrm{mem}}$. We do not detect any significant dependence of $\msbcg$ or bias ratio on $R_{\mathrm{mem}}$.}
\label{fig:rmem_split}
\end{center}
\end{figure*}

We first examine the distributions of the low and high-$\msbcg$ clusters on the
$R_{\mathrm{mem}}$ vs. $\lambda$ plane, which is shown on the left panel of
Figure~\ref{fig:rmem_split}. Red and blue contours indicate the $20\%$, $50\%$, and $90\%$
enclosed regions, while the red circles and blue squares with errorbars show the
median $R_{\mathrm{mem}}$ as functions of $\lambda$ for the high and
low-$\msbcg$ subsamples, respectively. The two sets of contours and median
relations are well aligned, showing no systematic offset between the high
and low-$\msbcg$ subsamples in $R_{\mathrm{mem}}$. The solid black line is a fit
to the median relations that we use to divide each $\msbcg$-based subsample into
low and high-$R_{\mathrm{mem}}$ quarter-samples for the test on the right panel.

The right panel of Figure~\ref{fig:rmem_split} is similar to the left panel of
Figure~\ref{fig:wpsignals_marked}, except for that we divide each of the low and
high-$\msbcg$ subsamples in half based on the solid black line in the left panel
and calculate the ratios between the high and low-$\msbcg$ signals within each
$R_{\mathrm{mem}}$ half in the bottom right panel. Filled and open red
circles~(blue squares) indicate the measurements for the high and
low-$R_{\mathrm{mem}}$ quarter-samples split from the high~(low)-$\msbcg$
subsample, respectively. The high-$R_{\mathrm{mem}}$ profiles exhibit enhanced
clustering on all scales above 400 $\hkpc$ than the low-$R_{\mathrm{mem}}$ ones
due to strong projection effects, echoing the findings in \citet{Sunayama2020}.
However, the amplitudes of the relative enhancement are the same between the low
and high-$\msbcg$ subsamples, indicating similar projection effects in the
high-$R_{\mathrm{mem}}$ clusters regardless of the BCG stellar mass.

Despite the strong projection effects of the high-$R_{\mathrm{mem}}$ clusters,
the ratio profiles in the bottom panels are both in good agreement with the
prediction from the conformity+AB model as in Figure~\ref{fig:wpsignals_marked},
though they are also consistent with the prediction from the conformity-only model
due to the large uncertainties.  Therefore, using the average
radius of the member galaxy candidates $R_{\mathrm{mem}}$ as a proxy of the
projection effect, we do not find any correlation between projection effect and
BCG stellar mass that could induce the strong conformity we detected among the
SDSS clusters, nor do we find any evidence that our detected cluster assembly bias
signal depends on the level of projection effects within the sample.

\section{Physical Implications}
\label{sec:physics}

\subsection{Could Dry Mergers Drive the Strong BCG-Satellite Conformity?}
\label{subsec:physics_merger}

The physical conformity between the BCG stellar mass and satellite richness
implies a correlated growth between the BCGs and satellite galaxies inside
clusters, and the increasing trend of $\rho_{\mathrm{cc}}$ with $\mh$ suggests
that the BCGs in the most massive haloes almost grow in lockstep with the
accretion of satellite galaxies. This strong conformity could naturally occur if
a significant fraction of the BCG stellar mass growth is {\it ex situ}, via the
dry mergers with massive satellite galaxies that were transported to the cluster
centre by dynamical friction~\citep{Chandrasekhar1943, White1976}.  Indeed,
observations indicate that the mode of BCG stellar mass growth switched from
{\it in situ} star formation to {\it ex situ} stellar accretion around
$z{\sim}1$~\citep{Webb2015, McDonald2016, Vulcani2016, Lavoie2016,
Groenewald2017, Zhao2017}.

In order for dry mergers to drive a correlated scatter between BCG stellar mass
and satellite richness, the merger-induced stellar growth should be significant,
e.g., comparable with the intrinsic scatter in the cluster
SHMR of ${\sim}0.05{-}0.1$ dex~\citep{Golden-Marx2021}.
Observationally, the stellar growth from dry mergers
since $z{\sim}1$ varies between $30\%$~\citep{Collins2009, Bundy2017, Lin2017} and almost a factor of two~\citep{Whiley2008, Burke2013, Lidman2013}. However,
semi-analytic models~(SAMs) predict that the BCG stellar mass could grow by a
factor of 3-4 between $z={1}$ and $z={0}$~\citep{DeLucia2007, RUSZKOWSKI2009,
Laporte2013, Oogi2016}. The discrepancy between SAM predictions and observations
is partly due to the numerical uncertainties in modelling dynamical
friction~\citep{Jiang2008, Boylan-Kolchin2008}, and it is also unclear what
fraction of the accreted stars would end up in the diffuse intra-cluster
light~\citep{Murante2007, Contini2018}.

Alternatively, using a self-consistent model of the observed conditional stellar
mass functions across cosmic time, \citet{Yang2013} carefully accounted for the
total amount of {\it in situ} growth by modelling the star formation histories
of central galaxies as a function of halo mass, stellar mass, and redshift.
Their indirect method estimated that at $z{\geq}2.5$
less than 1\% of the stars in the progenitors of massive galaxies are formed {\it ex
situ}, but this fraction increases rapidly with redshift, becoming ${\sim}40$\% at
$z{=}0$. Therefore, by combining the observations, SAM predictions, and indirect
estimates, we expect the average amount of merger-induced stellar mass growth to
be between $0.1-0.3$ dex, hence more than enough for driving a correlated scatter with
richness.

Finally, the observed richness roughly corresponds to the number of massive, quenched
satellite galaxies in each cluster, i.e., the same type of galaxies that would
preferentially merge with the BCG within a Hubble time. For instance,
\citet{Boylan-Kolchin2008} estimated that roughly $10{-}20\%$ of all the accreted
satellites with mass ratio above $1{:}10$ would merge with the BCG within 7
G$yrs$ due to dynamical friction~\citep[see also][]{Jiang2008}. As a result, the
observed conformity between BCG stellar mass and richness could be strongly boosted by
the fact that the low-mass, star-forming satellites are often not included when
calculating the richness of optical clusters.

\subsection{Is BCG-Satellite Conformity Consistent with the
\texorpdfstring{$\bm{c}-\bm{\msbcg}$}{c-Mstar} and \texorpdfstring{$\bm{c}-\bm{\lambda}$}{c-lambda} Relations?}
\label{subsec:physics_concentration}

\begin{figure*}
\begin{center}
    \includegraphics[width=0.96\textwidth]{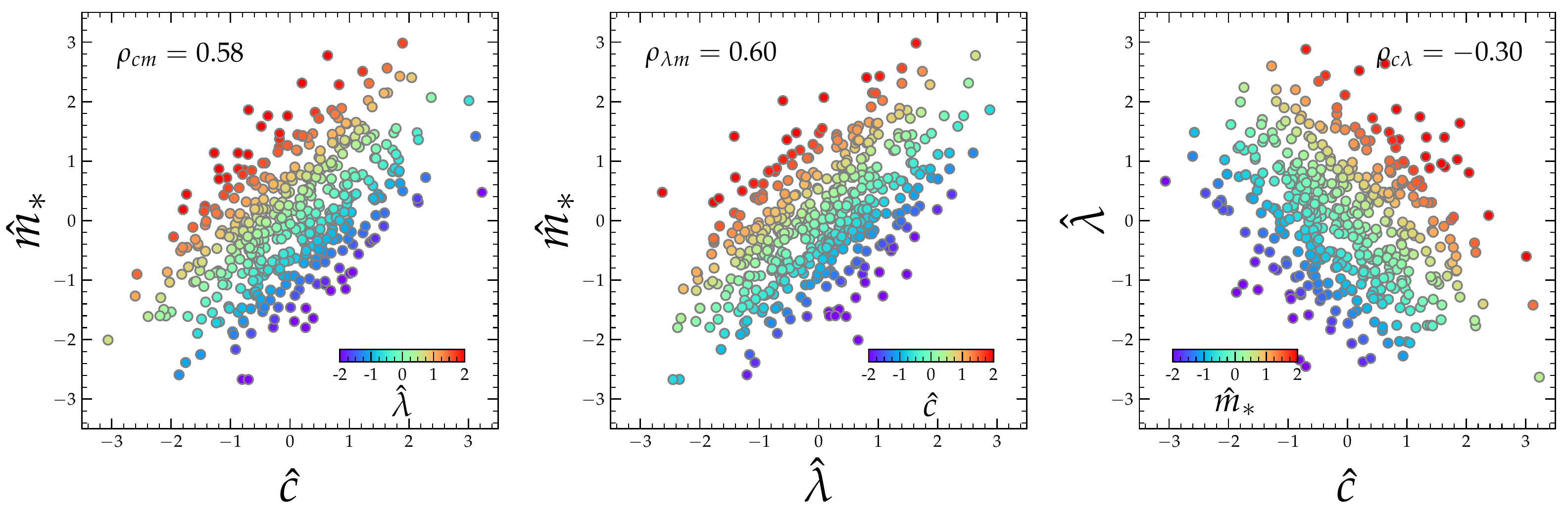} \caption{
	A toy model illustrating the relations between the relative BCG stellar
	mass $\hat{m}_*$, relative richness $\hat{\lambda}$, and relative
	concentration $\hat{c}$ at fixed halo mass. The three panels show the
	distributions of 500 mock clusters generated with
	Equation~\ref{eqn:toy} on the planes of $\hat{c}{-}\hat{m}_*$~(left),
	$\hat{\lambda}{-}\hat{m}_*$~(middle), and
	$\hat{c}{-}\hat{\lambda}$~(right), respectively. In each panel, each
	filled circle is colour-coded by the value of the third quantity~(other
	than the quantities in the x and y axes) according to the inset colourbar. The Pearson
	cross-correlation coefficient is shown by the legend on the top left of
	each panel. Our toy model of Equation~\ref{eqn:toy} successfully
	reproduces the observed correlation between concentration and BCG
	stellar mass~(left), BCG-satellite conformity~(middle), and the
	simulation-predicted anti-correlation between concentration and
	richness~(right).}
\label{fig:toymodel}
\end{center}
\end{figure*}

\citetalias{Zu2021} showed that halo concentration is one of the key drivers of
scatter in the SHMR of clusters, so that clusters with more concentrated cores
host more massive BCGs at fixed halo mass --- a positive $c{-}\msbcg$
correlation. In \citetalias{Zu2021}, we speculated
that the correlation between $c$ and $\msbcg$ is caused
by the fact that the {\it in situ} stellar mass growth of the BCGs is closely
tied to the rapid growth of dark matter mass at early times. At the onset of
cluster formation, fast accretion and frequent mergers not only built up the
central cores of dark matter haloes~\citep{Zhao2013, Klypin2016}, but also drove
strong starbursts in the progenitors of the BCGs via rapid cooling flows and
shocks, respectively~\citep{Fabian1994, McDonald2012, Barnes1991, Mihos1996,
Hopkins2013}.

Meanwhile, there exists a well-known anti-correlation between the concentration
and substructure abundance of haloes at fixed halo mass~\citep{Giocoli2010},
which should translate to a $c{-}\lambda$ anti-correlation at fixed $\mh$.
Combining this anti-correlation with the positive correlation between $c$ and
$\msbcg$, one might naively expect that the clusters with high $\lambda$~(hence
low $c$) would host less massive BCGs than their low-$\lambda$ counterparts at
fixed $\mh$ --- an anti-conformity between $\msbcg$ and $\lambda$, in apparent
contradiction with our finding of a strong BCG-satellite conformity from the
data.

Before delving into the astrophysics in \S\ref{subsec:physics_toy}, the expectation of a BCG-satellite
anti-conformity is a statistical fallacy, as correlations are non-transitive
properties --- the observed correlation between $\msbcg$ and $c$, combined with
the anti-correlation between $c$ and $\lambda$, does not necessarily yield a
negative correlation between $\msbcg$ and $\lambda$. For the correlation matrix
between the three quantities $c$, $\lambda$, and $\msbcg$,
\begin{equation}
    \mathbfss{A} =
    \begin{pmatrix}
        1  & \rho_{\mathrm{c}\lambda} & \rho_{\mathrm{cm}}\\
        \rho_{\mathrm{c}\lambda} & 1 & \rho_{\lambda \mathrm{m}} \\
        \rho_{\mathrm{cm}} & \rho_{\lambda \mathrm{m}} & 1 \\
\end{pmatrix},
\end{equation}
where $\rho_{\mathrm{c}\lambda}$, $\rho_{\mathrm{cm}}$, and $\rho_{\lambda
\mathrm{m}}$ are the correlation coefficients between the three pairs of
quantities indicated by the subscripts~($\rho_{\lambda \mathrm{m}}$ is equivalent to our conformity parameter $\rho_{\mathrm{cc}}$ in~\S\ref{sec:method}). In order for $\mathbfss{A}$ to be
positive-definite, the determinant shall be positive,
\begin{equation}
    | \mathbfss{A} | = 1- \rho^2_{\lambda \mathrm{m}} - \rho^2_{\mathrm{c}\lambda}
    - \rho^2_{\mathrm{cm}} + 2\rho_{\mathrm{c}\lambda}\rho_{\lambda \mathrm{m}}
    \rho_{\mathrm{cm}} > 0,
    \label{eqn:det}
\end{equation}
so that $\rho_{\lambda \mathrm{m}}$ has to be in between
\begin{equation}
    \rho_{\mathrm{c}\lambda}\,\rho_{\mathrm{cm}}\pm
    \sqrt{\left(1-\rho^2_{\mathrm{c}\lambda}\right)\left(1-\rho^2_{\mathrm{cm}}\right)}.
\end{equation}
Assuming $\rho_{\mathrm{c}\lambda}{=}-0.3$ and $\rho_{\mathrm{cm}}{=}0.5$, we
obtain the statistically allowed range of $\rho_{\lambda \mathrm{m}}$ as
\begin{equation}
   -0.68 < \rho_{\lambda \mathrm{m}} < 0.98,
    \label{eqn:range}
\end{equation}
i.e., the correlation between BCG stellar mass and satellite richness can be as
strongly positive as possible given reasonable values of a negative
$\rho_{\mathrm{c}\lambda}$ and a positive $\rho_{\mathrm{cm}}$. For
$\rho_{\lambda \mathrm{m}}$ to be strictly below zero when the signs of $\rho_{\mathrm{c}\lambda}$ and $\rho_{\mathrm{cm}}$ are
different, we need
\begin{equation}
\rho^2_{\mathrm{c}\lambda} + \rho^2_{\mathrm{cm}} > 1.
\end{equation}
Therefore, for $\rho_{\lambda \mathrm{m}}$ to be strictly negative when
$\rho_{\mathrm{c}\lambda}{=}{-}0.3$, the correlation between concentration and BCG
stellar mass has to be greater than $\rho_{\mathrm{cm}}{=}0.95$ --- a plausible value but quite unlikely in the presence of {\it ex situ} BCG growth.

\subsection{A Toy Model for the \texorpdfstring{$\bm{\msbcg}-\bm{\lambda}-\bm{c}$}{Mstar-lambda-c} Relation at Fixed \texorpdfstring{$\bm{\mh}$}{Mh}}
\label{subsec:physics_toy}

The key to understanding the connection between $\msbcg$, $c$, and $\lambda$ at
fixed $\mh$ is to decompose the observed $\msbcg$ into two components of
different physical origins and formation epochs.  As discussed in
\S\ref{subsec:physics_merger}, the amount of {\it ex situ} BCG stellar mass is
likely related to the frequency of late-time BCG-satellite mergers, which is
directly tied to the number of massive quenched satellites, i.e., $\lambda$; The
{\it in situ} portion of $\msbcg$ is likely tied to $c$ due to the co-evolution
of BCGs and dark matter haloes in the early phase of cluster formation.
Therefore, it is plausible that at fixed $\mh$, $\msbcg$ is positively
correlated with {\it both} $c$ and $\lambda$, while $c$ and $\lambda$ are
themselves anti-correlated.

To illustrate such a physical connection between $\msbcg$, $c$, and $\lambda$ at
fixed $\mh$, we can build a simple toy model for explaining the scatter in
the BCG stellar mass using two separate components sourced by concentration and
richness. To remove the halo mass dependence in the toy model, we choose to
model the {\it relative} BCG stellar mass $\hat{m}_*$ defined in
Equation~\ref{eqn:mhat} using the {\it relative} concentration
$\hat{c}$~(Equation~\ref{eqn:chat}) and the {\it relative}
richness~(Equation~\ref{eqn:lhat}).

In particular, we assume $\hat{m}_*$ can be written as the sum of the {\it in situ} and {\it ex
situ} components with no extra source of stochastic variance,
\begin{equation}
    \hat{m}_* = \hat{m}_{*, \mathrm{in}} + \hat{m}_{*, \mathrm{ex}},
\end{equation}
where the two terms on the right-hand side can be described by
two linear terms of $\hat{c}$ and $\hat{\lambda}$, respectively,
\begin{equation}
    \hat{m}_* = f_{\mathrm{in}}\, \hat{c} + f_{\mathrm{ex}}\, \hat{\lambda},
    \label{eqn:toy}
\end{equation}
where
\begin{equation}
    f^2_{\mathrm{in}} + f^2_{\mathrm{ex}} + 2 \,f_{\mathrm{in}}\, f_{\mathrm{ex}}\, \rho_{\mathrm{c}\lambda} = 1.
    \label{eqn:var}
\end{equation}
The equality in Equation~\ref{eqn:var} is to ensure that $\hat{m}$ also has a
unit variance when both $\hat{c}$ and $\hat{\lambda}$ are unit-variance
Gaussians. Assuming a $\rho_{\mathrm{c}\lambda}$ of ${-}0.30$, we find
$f_{\mathrm{in}}{=}0.84$ and $f_{\mathrm{ex}}{\simeq}0.85$ so that the equality in
Equation~\ref{eqn:var} is reached and the resultant $\rho_{\lambda\mathrm{m}}$
is roughly the value we inferred at ${M_{\mathrm{pivot}}}$, i.e., $0.60$.

To produce a sample of mock clusters, we assume $\hat{c}$ and $\hat{\lambda}$
jointly follow a zero-means, unit-variances bivariate Gaussian with a correlation
coefficient of $-0.30$, and then generate $500$ random values of $\hat{c}$ and
$\hat{\lambda}$ from this anti-correlated 2D Gaussian. We then derive $500$
values of $\hat{m}_*$ using Equation~\ref{eqn:toy}.

Figure~\ref{fig:toymodel} shows the correlation for each of the three pairs of
quantities: $\hat{c}{-}\hat{m}_*$~(left), $\hat{\lambda}{-}\hat{m}_*$~(middle),
and $\hat{c}{-}\hat{\lambda}$~(right), respectively. In each panel, each filled
circle represents a mock cluster on the plane of the paired quantities,
colour-coded by the value of the third quantity, indicated by the horizontal
inset colourbar. The Pearson cross-correlation coefficient is indicated by
legend on the top left. Similar to the observations, $\hat{m}_*$ shows strong
positive correlations with both $\hat{c}$ and $\hat{m}$, with comparable
correlation coefficients of $0.59$ and $0.60$, respectively, despite that
$\hat{c}$ and $\hat{m}$ are by design negatively correlated. Therefore, such an
extremely simple model of Equation~\ref{eqn:toy} can qualitatively reproduce
the two key observations in \citetalias{Zu2021} and in this paper, the
BCG-concentration correlation~(left) and the BCG-satellite conformity~(middle),
respectively, without breaking the concentration-richness anti-correlation
robustly predicted by simulations~(right). The success of this toy model is very
encouraging, pointing at a viable path to building a more comprehensive model of
$\msbcg-\lambda-c$ connection for future cluster surveys.

\section{Summary and Conclusion}
\label{sec:conc}

We have inferred the level of conformity within the SDSS redMaPPer clusters,
defined as the correlation coefficient $\rho_{\mathrm{cc}}$ between the BCG
stellar mass $\msbcg$ and satellite richness $\lambda$ at fixed halo mass, using
the observed abundance and weak lensing of clusters as functions of both
$\msbcg$ and $\lambda$. With the richness-halo mass relation largely anchored by
the weak lensing mass of clusters binned in richness
\begin{equation}
    \avg{\ln\lambda \mid \mh}  = 3.44 + 1.06\,\ln\left(\mh/3\times10^{14}\right),
\end{equation}
our best--fitting conformity model with
\begin{equation}
    \rho_{\mathrm{cc}}(\mh)= 0.60 + 0.08\,\ln\left(\mh/3\times10^{14}\right)
\end{equation}
can successfully resolve the ``halo mass equality'' conundrum discovered
in~\citet{Zu2021} --- when split by $\msbcg$ at fixed $\lambda$, the low and
high-$\msbcg$ clusters have the same average weak lensing halo mass, despite the
$0.34$ dex discrepancy in their average BCG stellar mass.  Our method of
reconstructing the interconnection between multiple cluster observables using
the abundance and weak lensing of clusters can be naturally extended to X-ray
and SZ~(Sunyaev-Zel'dovich) surveys of clusters~\citep{Stanek2006, Miyatake2019,
Chiu2021, Nicola2020}.

We develop a prescription for the cluster assembly bias effect that ties the
halo concentration measured by small-scale $\ds$ to the cluster bias measured by
either $\ds$ or cluster-galaxy cross-correlation on large scales. By combining
cluster conformity with assembly bias, we build an accurate model for the weak
lensing profiles $\ds$ of the low and high-$\msbcg$ clusters across all distance
scales. Our conformity+AB model of $\ds$ predicts that the high-$\msbcg$
clusters have ${\sim}20\%$ more concentrated~($c{=}6.95$) dark matter haloes, but are ${\sim}10\%$ less
biased~($b{=}2.85$) than the low-$\msbcg$ clusters~($c{=}5.87$ and
$b{=}3.11$), in good agreement with the observations. Using the average
membership distance as a proxy of the background contamination, we
demonstrate that the impact of projection effects on the inferred
conformity and assembly bias signal is likely small~\citep{Zu2017, Busch2017, Sunayama2020}.

We argue that a simple picture of the two-phase BCG-halo co-evolution can
explain the complex connection between $\msbcg$, $\lambda$, and $c$ at fixed
halo mass, i.e., $\msbcg$ is positively correlated with both $c$ and $\lambda$
despite the anti-correlation between $c$ and $\lambda$. In this simple picture,
the starbursting phase of the BCG {\it in situ} growth is induced by the rapid
accretion and frequent mergers that built up the central core of the cluster
haloes at high redshift, while the {\it ex situ} BCG stellar mass growth at late
times is predominantly driven by the dry mergers with the massive satellites that
sunk into the cluster centres via dynamical friction. Consequently, the {\it in
situ} portion of $\msbcg$ is tied to the halo concentration, while the
{\it ex situ} portion of $\msbcg$ naturally correlates with the richness of
satellite galaxies. A simple toy model based on this physical picture can
qualitatively reproduces the salient features of the observed
$\msbcg$-$c$-$\lambda$ connection.

The strength of the inferred conformity signal may depend on the cluster finder,
especially the centroiding algorithm and the definition of richness. We plan to
extend our analysis to other publicly-available cluster catalogues, e.g., the
\citet{Yang2021} halo-based group catalogue from DECaLS imaging~\citep[see
also][]{Tinker2020, Zou2021} and the \citet{Wen2021} cluster catalogue based on
HSC and WISE.  Furthermore, the conformity signal could also depend on
cosmology.  \citet{Murata2019} showed that while the constraints on the mean
richness-halo mass relation are consistent between the {\it Planck} and {\it
WMAP} models, the best--fitting scatter for {\it Planck} is progressively larger
than the {\it WMAP} model for lower-mass haloes. However, the conformity signal
is primarily constrained by the dependence of average halo mass on $\msbcg$ at
fixed richness, therefore should be less affected by the size of the scatter in
the richness-halo mass relation.

With the ever-increasing precision of cluster weak lensing
measurements~\citep{Mandelbaum2018}, we will be able to routinely measure not
only the average halo mass of clusters, but also the average halo concentration
robustly from the shape of $\ds$ on small scales, after marginalising over the
mis-centring~\citep{Zhang2019} and baryonic effects~\citep{Cromer2021}.
Meanwhile, the diminishing statistical uncertainties of cluster surveys demand a
thorough physical understanding of the galaxy-halo connection at the high mass
end, which would greatly mitigate the systematic uncertainties in cluster
cosmology~\citep{Wu2019, Wu2021} via the making of more realistic synthetic clusters~\citep{Varga2021}.
More important, an observationally-motivated yet physically-comprehensive model
of galaxy-halo connection, e.g., an extension to our toy model of the
$\msbcg$-$c$-$\lambda$ connection in~\S\ref{subsec:physics_toy}, could point us
to a minimum-scatter proxy of halo mass~\citep{Palmese2020, Bradshaw2020,
Farahi2020, Tinker2021}.  Therefore, it is imperative that we incorporate the
strong conformity and assembly bias effect into the modelling of galaxy-halo
connection and weak lensing of clusters for next-generation cluster surveys,
including the Rubin Observatory Legacy Survey of Space and
Time~\citep[LSST;][]{Ivezic2019}, {\it Euclid}~\citep{Laureijs2011}, Chinese
Survey Space Telescope~\citep[{\it CSST};][]{Gong2019}, and the Roman Space
Telescope~\citep[{\it Roman};][]{Spergel2015}.

\section*{Data availability}

The data underlying this article will be shared on reasonable request to the corresponding author.

\section*{Acknowledgements}

We thank the anonymous referee for the helpful suggestions that have
greatly improved this manuscript. We thank Weiguang Cui, Melanie Simet,
and Rachel Mandelbaum for helpful discussions.  We gracefully thank
Christopher Conselice for suggesting the term ``cluster conformity'' for
describing the correlation between BCG and satellites.  YZ acknowledges the
support by the National Key Basic Research and Development Program of China
(No. 2018YFA0404504), National Science Foundation of China (11873038,
11621303, 11890692, 12173024), the science research grants from the China
Manned Space Project (No.  CMS-CSST-2021-A01, CMS-CSST-2021-B01), the
National One-Thousand Youth Talent Program of China, and the SJTU start-up
fund (No.  WF220407220). YZ and YPJ acknowledge the support by the 111
Project of the Ministry of Education under grant No. B20019. YZ thanks the
wonderful hospitality by Cathy Huang during his visit at the Zhangjiang
Hi-Tech Park during the summer of 2021.






\appendix

\section{A Prescription for Cluster Assembly Bias}
\label{sec:dsfull_cab}

\begin{figure*}
\begin{center}
    \includegraphics[width=0.96\textwidth]{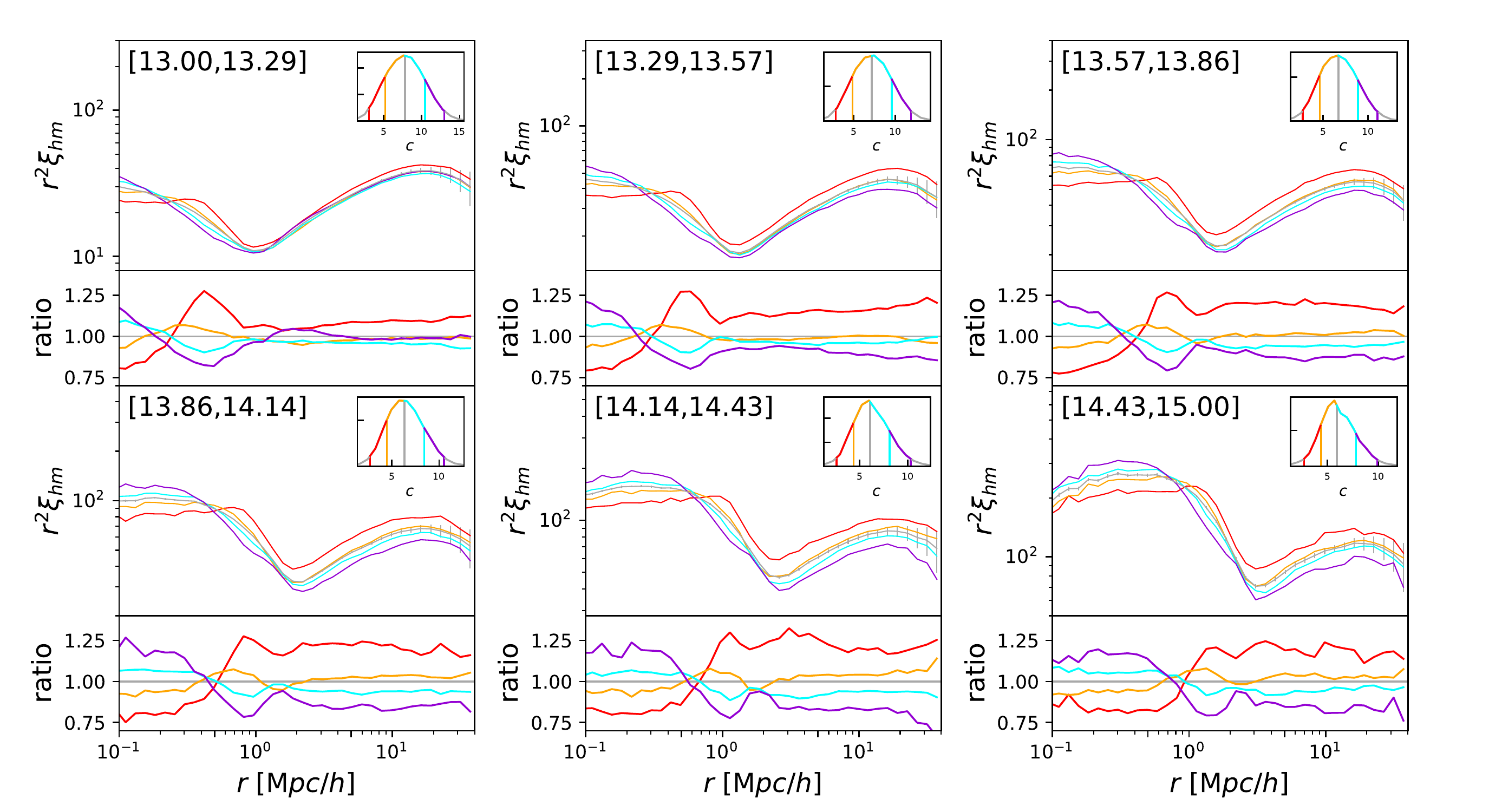}
    \caption{Dependence of the 3D isotropic halo-matter cross-correlation
    function $\xihm$ on concentration in six halo mass bins ranging from
    $10^{13}\hmsol$ to $10^{15}\hmsol$~(increasing from left to right and
    from top to bottom).  For each mass bin, the main sub-panel compares
    the $\xihm$ of haloes in four concentration bins, $[-2\sigma_c,
    \sigma_c]$~(red), $[-\sigma_c, 0]$~(orange), $[0, \sigma_c]$~(cyan),
    and $[\sigma_c, 2\sigma_c]$~(purple), illustrated by the four segments
    of concentration distribution in the inset panel. We plot $r^2\xihm$
    instead of $\xihm$ in the y-axes to highlight the difference between
    the four concentration bins on all distance scales.  Gray curve with
    errorbars indicates the $\xihm$ of all the haloes in that mass bin. The
    bottom sub-panel shows the ratio profiles between the $\xihm$ of four
    concentration bins and that of all haloes in the same mass bin. We
    calculate halo biases from the $\xihm$ measurements on scales between
    $10\hmpc$ and $30\hmpc$, where the biases are roughly linear.}
\label{fig:xihm}
\end{center}
\end{figure*}

\begin{figure}
\begin{center}
    \includegraphics[width=0.48\textwidth]{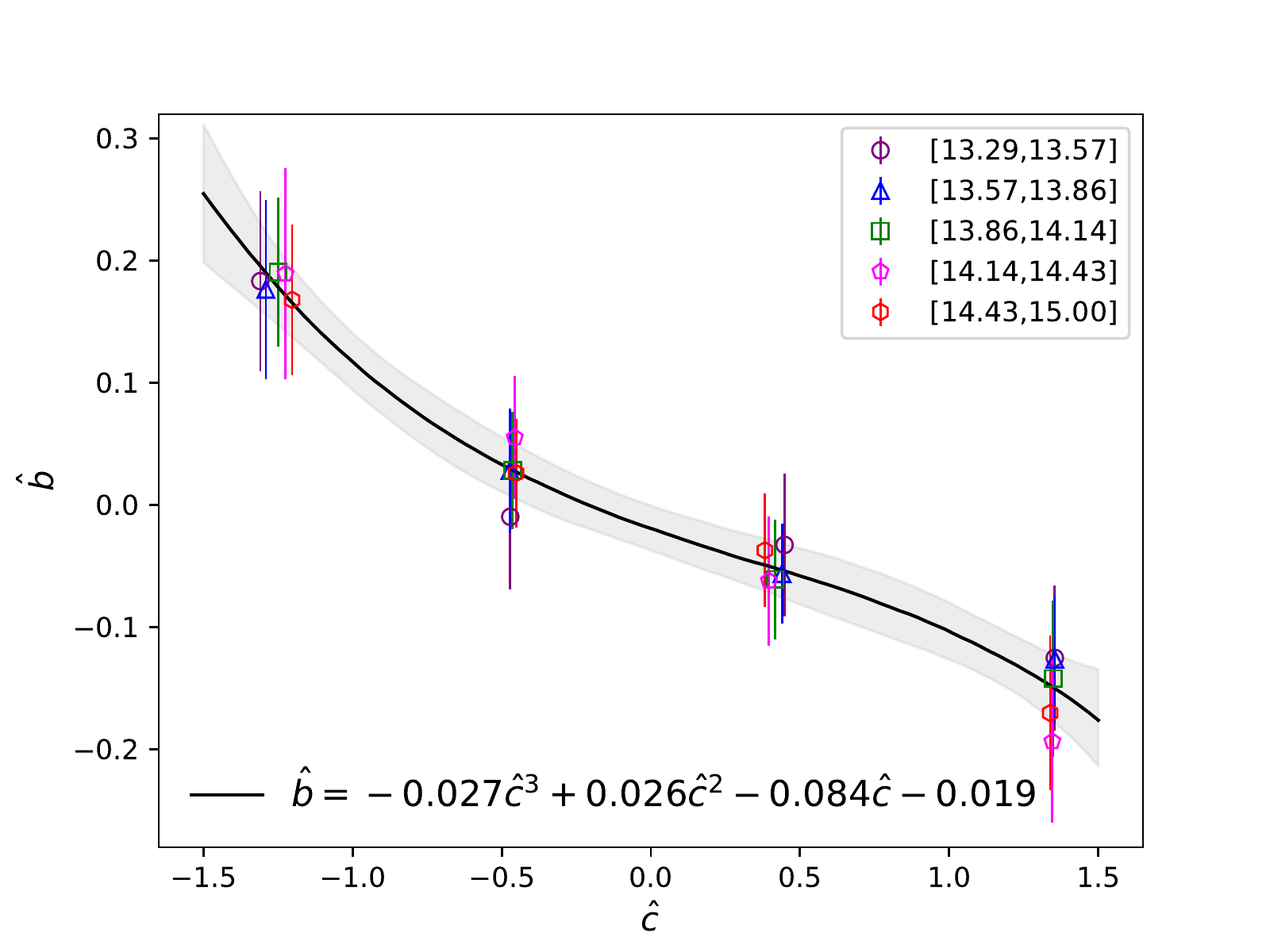} \caption{Our
    prescription of the cluster assembly bias calibrated for haloes at $z{\simeq}0.23$. Symbols of different colours with errorbars
    indicate the measurements of the relative bias
    $\hat{b}$~(Equation~\ref{eqn:bhat}) as a function of the relative
    concentration $\hat{c}$~(Equation~\ref{eqn:chat}) in five different halo
    mass bins indicated by the legend in the top right.  Black solid line is our
    3rd-order polynomial fit to the data points, with the best--fitting
    parameters indicated in the bottom left.}
\label{fig:hab}
\end{center}
\end{figure}

To calibrate an accurate prescription of cluster assembly bias, we employ a
large-volume high-resolution cosmological $N$-body simulation from the
\texttt{CosmicGrowth} suite developed by \citet{Jing2019}. In particular, we
utilize the $z{=}0.23$~(closest to the mean redshift of our cluster sample)
snapshot of the \texttt{Planck\_2048\_1200} simulation, which has a box-length
of $1.2\,\mathrm{G}pc/h$ and a mass resolution of $1.76\times10^{10}\hmsol$ at
{\it Planck} cosmology. We refer readers to \citet{Jing2019} for technical
details of the simulation. We identify dark matter haloes using the spherical
overdensity-based \texttt{ROCKSTAR}~\citep{Behroozi2013} halo finder, and
compute halo concentrations using the maximum circular velocity-based
approach~\citep{Klypin2011, Prada2012}.

We select all the haloes with mass between $10^{13}-10^{15}\hmsol$ and divide
them into six bins in halo mass. Within each halo mass bin, we measure the
median $\bar{c}$ and scatter $\sigma_c$ of the concentration distribution, and
select the haloes within $\pm 2\sigma_c$ into four concentration bins with equal
$1{-}\sigma_c$ widths. We then measure the 3D isotropic halo-matter
cross-correlation functions $\xihm$ by cross-correlating the positions of haloes
with that of dark matter particles, as shown in Figure~\ref{fig:xihm}.  The six
panels of Figure~\ref{fig:xihm} present the $\xihm$ measurements for the six
halo mass bins of $\lg\mh{=}13{-}13.29$, $13.29{-}13.57$, $13.57{-}13.86$,
$13.86{-}14.14$, $14.14{-}14.43$, and $14.43{-}15$, respectively. In each panel,
the main sub-panel shows the $\xihm$ of haloes in four concentration bins,
$[-2\sigma_c, \sigma_c]$~(red), $[-\sigma_c, 0]$~(orange), $[0,
\sigma_c]$~(cyan), and $[\sigma_c, 2\sigma_c]$~(purple), as well
as the measurement for all the haloes in that mass bin~(gray with errorbars).
The division of concentration bins is indicated by the concentration
distribution in the inset panel, with each coloured segment mapped to one of the
four concentration bins. We plot $r^2\xihm$ instead of $\xihm$ in the y-axis to
highlight the differences between the four concentration bins on both the small
and large scales. The bottom sub-panel shows the ratio between the $\xihm$
profile of each concentration bin and that of the all the haloes in that mass
bin. We compute the uncertainties of $\xihm$ and their ratios with Jackknife
re-sampling, though we do not show the errorbars~(except for the gray curves) in
Figure~\ref{fig:xihm} to avoid clutter.

The dependence of $\xihm$ on halo concentration is consistent across all mass
bins, with the low-concentration haloes showing stronger biases than the
high-concentration systems. The probability distributions of concentration are
reasonbaly Gaussian, with some level of skewness and kurtosis develped for the
higher mass bins. The lowest mass bin, however, does show a narrower range of
variation of bias with concentration, as the concentration-bias relation would
reverse its sign below the characteristic non-linear mass scale, i.e.,
low-concentration haloes would be less biased in the low halo mass
regime~\citep{Gao2005}. There, we drop the lowest mass bin of
Figure~\ref{fig:xihm} from our assembly bias calibration.  To accurately
calibrate assembly bias well into the characteristic non-linear mass scale, a
suite of extremely high-resolution simulations that can resolve haloes down to
$10^{11}\hmsol$ is required, hence beyond the scope of this paper.  Nonetheless,
since the main bulk of the halo mass distribution of our cluster sample is above
$\lg\mh{=}13.29$~(see Figure~\ref{fig:mstar_split}), the systematic uncertainty
of our assembly bias model caused by the omission of low mass haloes should be
small compared to the observational errors.

Given the similarities of assembly bias across the cluster mass range, we can
normalize the values of concentration and bias to remove the general trend of
$c$ and $b$ with halo mass. From each bin of ($\mh$, $c$) in
Figure~\ref{fig:xihm}, we can calculate the {\it relative} concentrations
$\hat{c}$ as
\begin{equation}
    \hat{c}(\mh, c) =\frac{c - \bar{c}(\mh)}{\sigma_c(\mh)},
    \label{eqn:chat}
\end{equation}
and the {\it relative} bias $\hat{b}(\mh, c)$ as
\begin{equation}
    \hat{b}(\mh, c) = \frac{b(\mh, c) - \bar{b}(\mh)}{\bar{b}(\mh)},
    \label{eqn:bhat}
\end{equation}
where $\bar{b}$ is the average halo bias of that halo mass. We calculate the
biases using the $\xihm$ measurements on scales between $10\hmpc$ and $30\hmpc$,
and the errorbars on $\hat{b}$ with Jackknife re-sampling technique.

Figure~\ref{fig:hab} shows our result of the cluster assembly bias measurement
in the form of $\hat{b}$--$\hat{c}$ relation in five different halo mass bins of
$13.29{-}13.57$, $13.57{-}13.86$, $13.86{-}14.14$, $14.14{-}14.43$, and
$14.43{-}15.00$, respectively. The relative bias exhibits a declining trend with
$\hat{c}$, reproducing the standard halo assembly bias phenomenon, i.e., an
anti-correlation between concentration and bias in the cluster mass
regime~\citep{Jing2007}. More important, the five $\hat{b}(\hat{c})$
measurements are consistent with each other, suggesting a universal
$\hat{b}$--$\hat{c}$ relation that is roughly independent of halo mass at
$\lg\mh{>}13.29$. As a result, we are able to fit a simple 3rd order polynomial
to the five mass bins simultaneously, yielding the black solid curve as our
prescription for the halo assembly bias in our cluster sample,
\begin{equation}
    \hat{b} = -0.027 \hat{c}^3 + 0.026 \hat{c}^2 - 0.084 \hat{c} - 0.019.
    \label{eqn:hab}
\end{equation}
Equation~\ref{eqn:hab} has a $\chi^2$ per degree of freedom of ${\sim}0.8$,
indicating a reasonably good description of the simulation measurements. Note that Equation~\ref{eqn:hab} does not go through $(0, 0)$, due to the fact that the concentration distribution at fixed $\mh$ is often slightly skewed.
Although our prescription is calibrated at {\it Planck} cosmology, it is likely
that the same parameters would still work for other
cosmologies~\citep{Contreras2021, Lazeyras2021}.

\section{A Joint Model of Conformity and Assembly Bias For \texorpdfstring{$\bm{\ds}$}{DelSig}}
\label{sec:dsfull_joint}

We are now ready to construct a comprehensive model of the cluster weak
lensing profile $\ds$ by incorporating both the BCG-satellite conformity
constrained in~\S\ref{subsec:infer_constraint} and the cluster assembly
bias calibrated in Figure~\ref{fig:hab}. In particular, in
\S\ref{sec:dsfull} we demonstrate the efficacy of our $\ds$ models by
comparing their predicted full weak lensing profiles of the low and
high-$\msbcg$ subsamples to the measurements from~\citetalias{Zu2021}. We
predict $\ds$ for the two cluster subsamples in two separate steps, with
cluster assembly bias modelled in the first step and the BCG-satellite
conformity modelled in the second. We describe each step in turn below.

In the first step, we predict $\ds$ as a function of the projected distance
$r_p$ for haloes at fixed mass $\mh$ and concentration $c$,
\begin{equation}
    \ds(r_p \mid \mh, c) = \overline{\Sigma}({<}r_p \mid \mh, c) - \Sigma(r_p \mid \mh, c),
\end{equation}
where $\overline{\Sigma}({<}r_p | \mh, c)$ and $\Sigma(r_p | \mh, c)$ are
the average surface matter density interior to and at radius $r_p$,
respectively. In the absence of mis-centring, $\Sigma(r_p)$ can be
predicted by integrating the 3D isotropic halo-mass cross-correlation
function $\xihm(r)$ over the line of sight distance $r_{\pi}$,
\begin{equation}
    \Sigma(r_p\mid \mh, c) = \rho_{m} \int_{-\infty}^{+\infty}
    \!\!\xi_{\mathrm{hm}}\left(\sqrt{r_p^2+r_{\pi}^2}\mid \mh, c\right)\;\mathrm{d} r_{\pi},
\end{equation}
where $\rho_{m}$ is the mean density of the Universe and we use
$\pm\,100\,\hmpc$ for the integration limit instead of $\pm\infty$ in practice.

Following~\citetalias{Zu2021}, we describe the mis-centring effect using the fraction of BCGs mis-centred
$f_{\mathrm{off}}$ and their offsets from the true centres
$r_{\mathrm{off}}$, which follows a shape-2 Gamma distribution
$p(r_{\mathrm{off}})$ with a characteristic offset $\sigma_{\mathrm{off}}$,
\begin{equation}
    p(r_{\mathrm{off}}) =  \frac{r_{\mathrm{off}}}{\sigma_{\mathrm{off}}^2}
    \exp\left(-\frac{r_{\mathrm{off}}}{\sigma_{\mathrm{off}}}\right).
\end{equation}
The observed surface matter density in the presence of mis-centring is thus
\begin{equation}
    \Sigma^{\mathrm{obs}}(r_p\mid \mh, c) =
    f_{\mathrm{off}}\,\Sigma^{\mathrm{off}}(r_p\mid \mh, c)\; +\; (1 -
    f_{\mathrm{off}})\,\Sigma(r_p\mid \mh, c),
\end{equation}
where
\begin{equation}
    \Sigma^{\mathrm{off}}(r_p) = \frac{1}{2\pi}
    \int_0^{\infty}\!\!\!\!\!\mathrm{d} r_{\mathrm{off}}\, p(r_{\mathrm{off}})
    \int_0^{2\pi} \!\!\!\!\!\mathrm{d}\theta\,\Sigma\left(\sqrt{r_p^2 + r_{\mathrm{off}}^2 - 2 r_p
    r_{\mathrm{off}} \cos\theta }\right).
\end{equation}
We adopt the best--fitting values of $f_{\mathrm{off}}$~(0.37 vs. 0.20 for low
and high-$\msbcg$ subsamples) and $\sigma_{\mathrm{off}}$~(0.23 vs. 0.21 $\hmpc$
for low and high) listed in the table 1 of~\citetalias{Zu2021}. For our current analysis, we
assume that both offset parameters are independent of halo mass for simplicity,
but expect to incorporate $\mh$--dependent mis-centring models for future
observations.

To calculate $\Sigma(r_p|\mh, c)$, we adopt the $\xi_{\mathrm{hm}}$ model developed by
\citet{Zu2014}~\citep[a modified version proposed by][]{Hayashi2008},
\begin{eqnarray}
\xihm(r\mid\mh, c) &=&   \left\{
\begin{array}{ll}
 \xi_\mrm{1h} & \quad\mbox{if $\xi_\mrm{1h} \geqslant \xi_\mrm{2h}$ },\nonumber\\
 \xi_\mrm{2h} & \quad\mbox{if $\xi_\mrm{1h} < \xi_\mrm{2h}$ },\nonumber
\end{array}
\right.\\
\xi_\mrm{1h} &=& \frac{\rho_\mrm{NFW}(r|M_h, c)}{\rho_\mrm{m}} - 1, \nonumber\\
    \xi_\mrm{2h} &=& b(\mh, c) \; \ximm.
\label{eqn:xihm}
\end{eqnarray}
Here $\xi_\mrm{1h}$ and $\xi_\mrm{2h}$ are the so-called ``1-halo'' and
``2-halo'' terms in the halo model~\citep{Cooray2002},
$\rho_\mrm{NFW}(r|\mh, c)$ is the NFW density profile of a halo with mass
$M_h$ and concentration $c$, $b(\mh, c)$ is the large-scale bias of that
halo, and $\ximm$ is the non-linear matter-matter auto-correlation function
predicted at {\it Planck} cosmology~\citep{Takahashi2012}. \citet{Zu2014}
found that Equation~\ref{eqn:xihm} provides an adequate description of the
halo-matter cross-correlation functions measured from simulations, though
for future surveys it is more preferred to switch to an emulator-based
approach for predicting $\xihm$ for better accuracy~\citep{Salcedo2020}.

By applying our cluster assembly bias prescription
calibrated in \S\ref{sec:dsfull_cab}, we can accurately predict $b(\mh, c)$ as
\begin{equation}
    b(\mh, c) \equiv b(\mh, \hat{c}) = \bar{b}(\mh)\left(1+
    \hat{b}(\hat{c})\right),
\end{equation}
where $\hat{b}(\hat{c})$ is the assembly bias relation of
Equation~\ref{eqn:hab}.  We adopt the fitting formulae for the mean
concentration--mass $\bar{c}(\mh)$ and bias--mass $\bar{b}(\mh)$ relations
from \citet{Zhao2009} and \citet{Tinker2010}, respectively.

After obtaining the prediction for $\ds(r_p | \mh, c)$, in the second step we
derive the weak lensing profiles of the high and low-$\msbcg$ cluster
subsamples~(hereafter referred to as $\bm{S}_*^+$ and $\bm{S}_*^-$,
respectively) by integrating $\ds(r_p |\mh, c)$ over the underlying halo
distribution $p(\mh, c | \bm{S}_*^{\pm})$
\begin{equation}
    \ds(r_p|\bm{S}_*^{\pm}) = \iint\!\! \ds(r_p | \mh, c)\, p(\mh, c| \bm{S}_*^{\pm}) \; \dd c\, \dd \mh.
    \label{eqn:dsalmost}
\end{equation}
The 2D PDF $p(\mh, c | \bm{S}_*^{\pm})$ can be rewritten as
\begin{equation}
p(\mh, c| \bm{S}_*^{\pm}) = p(c | \mh, \bm{S}_*^{\pm})\, p(\mh | \bm{S}_*^{\pm}),
    \label{eqn:ps}
\end{equation}
where $p(\mh | \bm{S}_*^{\pm})$ can be derived from Equation~\ref{eqn:psum} and is
shown in the bottom~(top) right panel of Figure~\ref{fig:mstar_split} for each of the
two subsamples predicted by the (anti-)conformity model.

The concentration distribution $p(c | \mh, \bm{S}_*^{\pm})$, however, is very
challenging to infer from the current weak lensing measurements. We can
nonetheless simplify the problem as follows.~\citetalias{Zu2021} discovered that the
high-$\msbcg$ clusters have a higher average concentration, hence a higher
average relative concentration $\langle \hat{c} | \mh\rangle$, than their
low-$\msbcg$ counterparts. If we make a further {\it ansatz} that the
probability distribution of relative concentration $\hat{c}$ is the same at any
fixed halo mass for either subsample
\begin{equation}
    p(\hat{c} \mid \mh,\, \bm{S}_*^{\pm}) \equiv p(\hat{c} \mid \bm{S}_*^{\pm}),
\end{equation}
so that $\langle \hat{c} | \mh\rangle{\equiv}\langle \hat{c}\rangle$, then the
average concentrations of the high and low-$\msbcg$ clusters can then be
modelled by only two parameters, $\langle \hat{c}_{\mathrm{+}}\rangle$ and $\langle
\hat{c}_{\mathrm{-}}\rangle$, respectively. After switching the variable from $c$ to $\hat{c}$,
\begin{equation}
    p(c \mid \mh,\, \bm{S}_*^{\pm})\,\dd c = p(\hat{c} \mid \bm{S}_*^{\pm})\,\dd\hat{c},
    \label{eqn:pc}
\end{equation}
Equation~\ref{eqn:dsalmost} can be rewritten as
\begin{equation}
    \ds(r_p\mid\bm{S}_*^{\pm})=\iint\!\! \ds(r_p\mid \mh, \hat{c})\, p(\hat{c} |
    \bm{S}_*^{\pm}) \, p(\mh | \bm{S}_*^{\pm})\; \dd \hat{c}\, \dd \mh.
    \label{eqn:dspenul}
\end{equation}

Finally, after integrating out $p(\hat{c} | \bm{S}_*^{\pm})$ in
Equation~\ref{eqn:dspenul}, we arrive at our prediction for the stacked weak
lensing profile of subsample $\bm{S}_*^{\pm}$
\begin{equation}
    \ds(r_p\mid\bm{S}_*^{\pm})=\int_{M_h^{\mathrm{min}}}^{M_h^{\mathrm{max}}} \ds\left(r_p
    \mid \mh,\, \langle\hat{c}_{\pm}\rangle \right)\;  p(\mh \mid \bm{S}_*^{\pm})\; \dd \mh,
    \label{eqn:dsfinal}
\end{equation}
where the cluster assembly bias and BCG-satellite conformity are incorporated
into the first and second terms of the integrand, respectively.

Since the halo mass distributions $p(\mh \mid \bm{S}_*^{\pm})$ can be predicted
by Equation~\ref{eqn:psum} for any level of conformity, the only unknown piece in Equation~\ref{eqn:dsfinal} is
$\ds\left(r_p | \mh,\, \langle\hat{c}_{\pm}\rangle \right)$, which we predict as
follows. Given a value of $\langle\hat{c}_{\pm}\rangle$ for subsample
$\bm{S}_*^{\pm}$, we can compute the average concentration at fixed halo mass as
\begin{equation}
    \langle c \mid \mh,\, \bm{S}_*^{\pm}\rangle = \langle\hat{c}_{\pm}\rangle \sigma_c + \bar{c}(\mh),
\end{equation}
and the average bias at fixed halo mass as
\begin{equation}
    \langle b \mid \mh,\, \bm{S}_*^{\pm}\rangle = \bar{b}(\mh)\left(1 + \hat{b}(\langle\hat{c}_{\pm}\rangle)
    \right),
\end{equation}
respectively, where $\hat{b}(\langle\hat{c}_{\pm}\rangle)$ is the cluster
assembly bias we derived in Equation~\ref{eqn:hab}.

To obtain a good description of the measured cluster weak lensing profiles on
small scales, we adjust the values of $\avg{\hat{c}_\pm}$ so that the average
concentration of each subsample is consistent with the best--fitting average
halo concentrations derived in~\citetalias{Zu2021}, i.e., $\avg{c_-}{=}5.87$ for
the low-$\msbcg$ and $\avg{c_+}{=}6.95$ for the high-$\msbcg$ subsamples,
respectively. Assuming the mean concentration--mass relation
from~\citet{Zhao2009} and a constant concentration scatter of $\sigma_c{=}0.15$,
our best--fitting values are $\avg{\hat{c}_-}{=}0.156$ and
$\avg{\hat{c}_+}{=}1.045$ within the posterior mean conformity model, while in
the anti-conformity model they become $\avg{\hat{c}_-}{=}0.17$ and
$\avg{\hat{c}_+}{=}1.206$. By applying those values of $\avg{c_{\pm}}$, we can
further infer the bias values via the cluster assembly bias prescription of
Equation~\ref{eqn:hab}. In particular, within the conformity model, the biases are
$\avg{b_+}{=}3.20$ and $\avg{b_-}{=}3.21$ when cluster assembly bias is switched
off, and $\avg{b_+}{=}2.85$ and $\avg{b_-}{=}3.11$ when cluster Assembly bias is
on, respectively. We also infer $\avg{b_+}{=}3.78$ and $\avg{b_-}{=}2.79$ for
the anti-conformity but without applying cluster assembly bias, because the
model is already ruled out regardless of the existence of cluster assembly bias.
We adopt those best--fitting values of $\avg{\hat{c}_{\pm}}$ and $\avg{b_{\pm}}$ and
predict the $\ds$ profiles for the low and high-$\msbcg$ subsamples under
different assumptions of conformity and assembly bias.

\bsp	
\label{lastpage}

\end{document}